\documentclass[10pt,prd,preprintnumbers,amsmath,amssymb,nofootinbib]{revtex4}
\usepackage{graphics}
\usepackage{bm}
\usepackage{cancel}
\usepackage{color}
\usepackage{graphicx}
\usepackage{multirow}
\usepackage{amsmath}
\usepackage{mathrsfs}
\usepackage{diagbox}
\usepackage{slashed}
\usepackage{amsmath}
\usepackage{float}

\usepackage{array}

\usepackage{extarrows}
\usepackage{amssymb}
\usepackage{color}
\usepackage{xcolor}

\makeatletter
\newcommand{\thickhline}{\noalign {\ifnum 0=`}\fi \hrule height 1pt\futurelet \reserved@a \@xhline}
\newcolumntype{"}{@{\hskip\tabcolsep\vrule width 1pt\hskip\tabcolsep}}                             
\makeatother

\usepackage{graphicx} 

\usepackage[colorlinks,
            citecolor=blue,
            anchorcolor=red,
            menucolor=red,
            linkcolor=red,
            filecolor=red,
            runcolor=red,
            urlcolor=blue,
            frenchlinks=red]{hyperref}

\begin{document}

\title{Spectrum and rearrangement decays of tetraquark states with four different flavors}

\author{Jian-Bo Cheng$^1$, Shi-Yuan Li$^1$, Yan-Rui Liu$^1$, Yu-Nan Liu$^1$, Zong-Guo Si$^1$, Tao Yao$^1$}
\affiliation{$^1$School of Physics, Shandong University, Jinan, Shandong, 250100, China}

\date{\today}
\begin{abstract}
We have systematically investigated the mass spectrum and rearrangement decay properties of the exotic tetraquark states with four different flavors using a color-magnetic interaction model. Their masses are estimated by assuming that the $X(4140)$ is a $cs\bar{c}\bar{s}$ tetraquark state and their decay widths are obtained by assuming that the Hamiltonian for decay is a constant. According to the adopted method, we find that the most stable states are probably the isoscalar $bs\bar{u}\bar{d}$ and $cs\bar{u}\bar{d}$ with $J^P=0^+$ and $1^+$. The width for most unstable tetraquarks is about tens of MeVs, but that for unstable $cu\bar{s}\bar{d}$ and $cs\bar{u}\bar{d}$ can be around 100 MeV. For the $X(5568)$, our method cannot give consistent mass and width if it is a $bu\bar{s}\bar{d}$ tetraquark state. For the $I(J^P)=0(0^+),0(1^+)$ double-heavy $T_{bc}=bc\bar{u}\bar{d}$ states, their widths can be several MeVs.

\end{abstract}
\maketitle

\section{INTRODUCTION}\label{sec1}

Since 2003, heavy quark exotic XYZ or $P_c$ states have been observed in almost every year \cite{Tanabashi:2018oca,Olsen:2014qna,Olsen:2017bmm,Yuan:2018inv}. It is impossible to assign all of them as conventional mesons or baryons in the traditional quark model. Their observation is consistent with the fact that QCD does not prohibit the existence of multiquark hadrons. To understand the natures of the exotic states, a large number of theoretical studies in the literature were performed, which include discussions about the hadron structures (molecules, compact multiquarks, kinematic effects, or other nonresonant interpretations), investigations on the production and decay properties, and developments of research methods
\cite{Liu:2013waa,Esposito:2014rxa,Chen:2016qju,Hosaka:2016pey,Richard:2016eis,Chen:2016spr,Lebed:2016hpi,Esposito:2016noz,Guo:2017jvc,Ali:2017jda,Karliner:2017qhf,Liu:2019zoy,Guo:2019twa}.

In these exotic states, the $X(5568)$ claimed by the D0 Collaboration \cite{D0:2016mwd,Abazov:2017poh} is special in that it is a $\bar{b}su\bar{d}$ or $\bar{b}sd\bar{u}$ meson with valence quarks of four different flavors. Unfortunately, its existence was not confirmed by latter experimental analyses from LHCb \cite{Aaij:2016iev}, CMS \cite{Sirunyan:2017ofq}, CDF \cite{Aaltonen:2017voc}, and ATLAS \cite{Aaboud:2018hgx}. On the theoretical side, various discussions \cite{Agaev:2016urs,Xiao:2016mho,Albaladejo:2016eps,Kang:2016zmv,Lang:2016jpk,Chen:2016ypj,Lu:2016kxm,Sun:2016tmz,Wang:2019eig,Ke:2018jql,Ke:2018stp,Agaev:2016mjb,Agaev:2016ijz,Agaev:2016ifn,Chen:2016mqt,Wang:2016mee,Wang:2016wkj,Zanetti:2016wjn,Dias:2016dme,Tang:2016pcf,Zhang:2017xwc,Albuquerque:2016nlw,Wang:2016tsi,Stancu:2016sfd,Liu:2016ogz,Lu:2016zhe,Esposito:2016itg,Ali:2016gdg,Chen:2016npt,Goerke:2016hxf,Liu:2016xly,Jin:2016cpv,Burns:2016gvy,Guo:2016nhb,Yang:2016sws,He:2016yhd,Liu:2017vsf,Mutuk:2019uez,Sungu:2019ybf} had tried to understand the nature of $X(5568)$. Most of investigations could not give a natural explanation for its low mass. Its production is also difficult to understand \cite{Jin:2016cpv}. However, the nonconfirmation of $X(5568)$ does not exclude the existence of $\bar{b}su\bar{d}$ or $\bar{b}sd\bar{u}$ tetraquark and its partner states. In this article, we are going to explore systematically the spectrum and decay properties of tetraquark mesons with four different flavors. There are totally nine systems we may consider,
$bc\bar{s}\bar{n}$, $bs\bar{c}\bar{n}$, $bn\bar{c}\bar{s}$, $bu\bar{c}\bar{d}$ or $bd\bar{c}\bar{u}$, $bu\bar{s}\bar{d}$ or $bd\bar{s}\bar{u}$, $cu\bar{s}\bar{d}$ or $cd\bar{s}\bar{u}$, $bc\bar{u}\bar{d}$, $bs\bar{u}\bar{d}$, and $cs\bar{u}\bar{d}$, where $n=u$ or $d$. All of them are heavy quark states. Recent studies of such systems within various approaches can be found in the literature \cite{Chen:2013aba,Chen:2017rhl,Yu:2017pmn,Luo:2017eub,Agaev:2016dsg,Agaev:2017oay,Chen:2018hts,Wu:2018xdi,Lucha:2018dzq,Huang:2019otd,Wang:2019ieu}. One may also consult Refs. \cite{Chen:2016qju,Chen:2016spr,Liu:2019zoy} for more investigations.

Previously in Ref. \cite{Liu:2016ogz}, we presented a study of the spectra of $qq\bar{q}\bar{Q}$ ($Q=c,b$, $q=u,d,s$) systems in the color-magnetic interaction (CMI) model. One may find the concepts and methods of the model in Ref. \cite{Maiani:2004vq} where the heavy-light diquark was pioneeringly adopted in understanding the tetraquark nature of the $X(3872)$. From our calculation with the CMI model \cite{Liu:2016ogz}, it was found that the $X(5568)$ can be assigned as a tetraquark state if small quark masses are adopted. A number of its stable partner tetraquarks may also exist. However, the conclusion based on the original form of model Hamiltonian is sensitive to the quark masses and thus different conclusions may be reached. From the following studies of various systems within the same model \cite{Wu:2016gas,Chen:2016ont,Wu:2017weo,Luo:2017eub,Zhou:2018pcv,Wu:2016vtq,Li:2018vhp,Wu:2018xdi}, the multiquark masses estimated with the quark masses $m_n\approx362$ MeV ($n=u,d$), $m_s\approx540$ MeV, $m_c\approx1725$ MeV, and $m_b\approx5053$ MeV can be treated as upper limits. These values are extracted from the masses of the lowest-lying conventional hadrons. 
A more reasonable multiquark mass seems to be close to that determined with a related hadron-hadron channel whose threshold can give higher multiquark masses. For example, the masses of the recently observed $P_c$ states can be reproduced with this estimation method \cite{Aaij:2019vzc,Cheng:2019obk}. The present study will include an updated analysis on the spectra of some $qq\bar{q}\bar{Q}$ systems. Then one may get a deeper understanding of the $X(5568)$ state.

Besides the spectroscopy, the decay properties of hadrons are also important in understanding their internal structures. For the decays of multiquark states, the rearrangement mechanism plays a significant role. In Ref. \cite{Cheng:2019obk}, we tried to understand the decay properties of the newly observed $P_c(4312)$, $P_c(4440)$, and $P_c(4457)$ by the LHCb Collaboration \cite{Aaij:2019vzc} with a very simple model $H_{decay}=const$. We found surprisingly that it works in the $uudc\bar{c}$ case. Here we also discuss the possible rearrangement decays of the tetraquarks with four different flavors and explore what happens if we use the same method although the model is very crude. The decay properties of four-quark states studied with interpolating currents through Fierz rearrangement can be found in Refs. \cite{Chen:2019wjd,Chen:2019eeq}.

This article is organized as follows. In Sec. \ref{sec2}, we introduce the formalism to study the spectrum and decay widths in the CMI model. Then in Sec. \ref{sec3}, the selection of parameters and numerical results for various systems will be shown. The last section gives some discussions and the summary of the present work.

\section{Formulation}\label{sec2}

In the CMI model, we first consider the mass splittings of the $S$-wave tetraquark states by using the color-spin interaction term of the one-gluon-exchange potential and then we estimate the tetraquark masses by adding effective quark masses. One does not need to solve the bound state problem in the model since the contributions from the spacial part of the total wave function have been encoded in the effective parameters. In this section, what we need is to construct the flavor-spin-color wavefunction bases, diagonalize the CMI Hamiltonian, and choose appropriate estimation method for tetraquark masses. A simple decay model is also presented.

\subsection{Wave functions}

In this study, we do not consider the isospin breaking effects and the nine systems we will consider are defined as
\begin{eqnarray}\label{Flavor}
 F_{1-9}=
bc\bar{s}\bar{n},\ bs\bar{c}\bar{n},\ bn\bar{c}\bar{s},\ bu\bar{c}\bar{d},\ bc\bar{u}\bar{d},\ bu\bar{s}\bar{d},\ bs\bar{u}\bar{d},\ cu\bar{s}\bar{d},\ cs\bar{u}\bar{d}.
\end{eqnarray}
Note that the states containing two $n$'s have two isospin possibilities, $I=1$ and $I=0$. For example, the flavor wave function of $F_5$ is $(|bc\bar{u}\bar{d}\rangle+|bc\bar{d}\bar{u}\rangle)/\sqrt{2}$ if $I=1$ while it is $(|bc\bar{u}\bar{d}\rangle-|bc\bar{d}\bar{u}\rangle)/\sqrt{2}$ if $I=0$. For the spin$\otimes$color wave functions, we define
\begin{eqnarray}\label{Base Vectors}
J=0: && B_1=[(q_1q_2)^0_6(\bar{q}_3\bar{q}_4)^0_{\bar 6}]^0_1,\quad B_2=[(q_1q_2)^1_6(\bar{q}_3\bar{q}_4)^1_{\bar 6}]^0_1, \quad
B_3=[(q_1q_2)^0_{\bar 3}(\bar{q}_3\bar{q}_4)^0_3]^0_1,\quad B_4=[(q_1q_2)^1_{\bar 3}(\bar{q}_3\bar{q}_4)^1_3]^0_1,\nonumber\\
J=1:&&B_5=[(q_1q_2)^1_6(\bar{q}_3\bar{q}_4)^0_{\bar 6}]^1_1,\quad B_6=[(q_1q_2)^0_6(\bar{q}_3\bar{q}_4)^1_{\bar 6}]^1_1,\quad B_7=[(q_1q_2)^1_6(\bar{q}_3\bar{q}_4)^1_{\bar 6}]^1_1, \nonumber\\
&&B_8=[(q_1q_2)^1_{\bar 3}(\bar{q}_3\bar{q}_4)^0_3]^1_1,\quad B_9=[(q_1q_2)^0_{\bar 3}(\bar{q}_3\bar{q}_4)^1_3]^1_1,\quad B_{10}=[(q_1q_2)^1_{\bar 3}(\bar{q}_3\bar{q}_4)^1_3]^1_1, \nonumber \\
J=2: && B_{11}=[(q_1q_2)^1_6(\bar{q}_3\bar{q}_4)^1_{\bar 6}]^2_1,\quad B_{12}=[(q_1q_2)_{\bar 3}^1(\bar{q}_3\bar{q}_4)^1_3]^2_1.
\end{eqnarray}
Here, the superscripts (subscripts) represent the spins (color representations) of the quark-quark state, the antiquark-antiquark state, and the tetraquark state. The mixing effects between states with the same total spin will be considered.

When combining the flavor, color, and spin wave functions, one needs to include the constraints from the Pauli principle for the $F_5=bc\bar{u}\bar{d}$, $F_7=bs\bar{u}\bar{d}$, and $F_9=cs\bar{u}\bar{d}$ cases, but not for other cases. We tabulate the needed spin$\otimes$color base states for $F_5$, $F_7$, and $F_9$ in various $(I,J)$ combinations in Table \ref{IJ}.
\begin{table}[htbp]
\caption{Needed spin$\otimes$color base states for $F_5=bc\bar{u}\bar{d}$, $F_7=bs\bar{u}\bar{d}$, and $F_9=cs\bar{u}\bar{d}$ in various $(I,J)$ combinations.}\label{IJ}
\centering
\begin{tabular}{c|c|c|c}\hline
   &$J=0$&$J=1$&$J=2$\\
$I=1$&$B_1$, $B_4$&$B_5$, $B_9$, $B_{10}$&$B_{12}$\\
$I=0$&$B_2$, $B_3$&$B_6$, $B_7$, $B_8$   &$B_{11}$\\ \hline
\end{tabular}
\end{table}

\subsection{Color-magnetic interaction and mass estimation}

With the constructed wave functions, the mass splittings between states with different spins for a given system can be calculated by using the CMI Hamiltonian,
\begin{equation}\label{CMI}
\widehat{H}_{CMI}=-\sum_{i<j}C_{ij}(\vec{\sigma}_i\cdot\vec{\sigma}_j)(\widetilde{\lambda}_i\cdot\widetilde{\lambda}_j).
\end{equation}
Here, $C_{ij}$ is the effective coupling strength between the $i$th and $j$th quark components, $\vec{\sigma}_i$ is the Pauli matrix acting on the $i$th quark component, and $\widetilde{\lambda}_i=\lambda_i$ ($-\lambda^*_i$) for quarks (antiquarks) with $\lambda$ being the Gell-Mann matrix.

For the systems $F_{1,2,3,4,6,8}$, the three $\langle{\widehat{H}}_{\!C\!M\!I}\rangle$ matrices corresponding to the total spin $J=2$, 0, and 1 are respectively
\begin{eqnarray}
\left[\begin{array}{cc}
 \frac{2(5\alpha-2\tau)}{3} & -2\sqrt{2} \nu \\
 -2 \sqrt{2} \nu  & \frac{4(2\tau+\alpha)}{3}
\end{array}\right]_{J=2}, \quad
\left[\begin{array}{cccc}
 4 \tau                   & -\frac{10 \nu}{\sqrt{3}}   & 0                     & 2\sqrt{6} \alpha \\
 -\frac{10 \nu}{\sqrt{3}} & -\frac{4(\tau+5\alpha)}{3} & 2\sqrt{6}\alpha       & 4\sqrt{2} \nu \\
 0                        & 2\sqrt{6} \alpha           & -8\tau                & -\frac{4 \nu}{\sqrt{3}} \\
 2\sqrt{6}\alpha          & 4\sqrt{2} \nu              & -\frac{4\nu}{\sqrt{3}}& \frac{8(\tau-\alpha)}{3}\end{array}\right]_{J=0},\nonumber
 \end{eqnarray}
 \begin{eqnarray}\label{CMI123}
\left[\begin{array}{cccccc}
\frac{4(\tau-2\theta)}{3}       & \frac{10 \nu}{3}       &\frac{10\sqrt{2}\beta}{3} & 0                  &-2\sqrt{2} \alpha      & -4\mu \\
\frac{10 \nu}{3}       & \frac{4(2\theta+\tau)}{3}        &-\frac{10\sqrt{2}\mu}{3}& -2\sqrt{2} \alpha   &0                   & 4 \beta \\
\frac{10\sqrt{2}\beta}{3}&-\frac{10\sqrt{2}\mu}{3}&-\frac{2(2\tau+5\alpha) }{3}       & -4\mu            &4\beta             & 2\sqrt{2}\nu \\
0                    & -2 \sqrt{2} \alpha        &-4\mu                   & \frac{8(2\theta-\tau)}{3}      &\frac{4 \nu}{3}        &\frac{4\sqrt{2}\beta}{3}\\
-2 \sqrt{2} \alpha        & 0                    &4\beta                  & \frac{4 \nu}{3}      &-\frac{8(2\theta+\tau)}{3}       &-\frac{4\sqrt{2}\mu}{3}\\
-4 \mu                 & 4 \beta                  &2\sqrt{2} \nu           &\frac{4\sqrt{2}\beta}{3}&-\frac{4\sqrt{2}\mu}{3}&\frac{4(2\tau-\alpha)}{3}
\end{array}\right]_{J=1}.
\end{eqnarray}
Their respective base vectors are $(B_{11},B_{12})^T$, $(B_1,B_2,B_3,B_4)^T$, and $(B_5,B_6,B_7,B_8,B_9,B_{10})^T$. Here, we have defined several variables,
\begin{eqnarray}
&&\tau=C_{12}+C_{34},\quad \theta=C_{12}-C_{34},\nonumber\\
&&\alpha=C_{13}+C_{14}+C_{23}+C_{24},\nonumber\\
&&\beta=C_{13}-C_{14}+C_{23}-C_{24},\nonumber\\
&&\mu=C_{13}+C_{14}-C_{23}-C_{24},\nonumber\\
&&\nu=C_{13}-C_{14}-C_{23}+C_{24}.
\end{eqnarray}
These matrices are the same as those given in Ref. \cite{Wu:2018xdi} if one takes the same order of base states. For the remaining three systems $F_5$, $F_7$, and $F_9$, their CMI matrices can be read out from Eq. \eqref{CMI123} by noting that the corresponding base vectors (Table \ref{IJ}) are $B_{12}$, $(B_1,B_4)^T$, and $(B_5,B_9,B_{10})^T$ for the case $I=1$ and $B_{11}$, $(B_2,B_3)^T$, and $(B_6,B_7,B_8)^T$, for the case $I=0$.

After diagonalizing the matrices, one gets the eigenvalues of the CMI Hamiltonians. If we use $E_{CMI}$ to denote an eigenvalue, the tetraquark masses can be estimated with the mass formula
\begin{eqnarray}\label{mass-ori}
M=\sum_{i=1}m_i+E_{CMI},
\end{eqnarray}
where the effective quark mass $m_i$ absorbs contributions from quark kinetic energy, color-electric interactions, and color confinement. Since the parameters ($m_i$, $C_{ij}$) vary for different systems, the model cannot give accurate hadron masses. Large uncertainty is mainly caused by the values of the effective quark masses. In the multiquark studies, to reduce the uncertainty, one may use an alternative mass formula \cite{Liu:2019zoy}
\begin{eqnarray}\label{mass-ref}
M=[M_{ref}-(E_{CMI})_{ref}]+E_{CMI},
\end{eqnarray}
where a reference system is introduced. The formula means that we are calculating the mass splitting between the multiquark state and the reference system. It also means that the uncertainty of $\sum_{i=1}m_i$ in the considered state is replaced by the uncertainty of $\sum_{i=1}m_i$ in the reference state. A natural choice of the reference system is a hadron-hadron channel whose quark content is the same as the considered multiquark state. This leads to a question how to select the channel. From our investigations \cite{Wu:2016gas,Chen:2016ont,Wu:2017weo,Luo:2017eub,Zhou:2018pcv,Wu:2016vtq,Li:2018vhp,Wu:2018xdi,Cheng:2019obk}, it seems that a channel resulting in higher multiquark masses is more appropriate. Based on this assumption, we further tried to use the $X(4140)$ \cite{Aaltonen:2009tz,Chatrchyan:2013dma,Abazov:2013xda,Abazov:2015sxa,Aaij:2016nsc,Aaij:2016iza} as a reference state in Ref. \cite{Wu:2018xdi} by assuming it to be the lowest $1^{++}$ $cs\bar{c}\bar{s}$ compact tetraquark \cite{Stancu:2009ka,Wu:2016gas}. Then the uncertainty problem in $\sum_{i=1}m_i$ becomes the uncertainty problem in mass differences between different quark flavors, e.g. uncertainty in $\Delta_{sn}=m_s-m_n$. If we use $\Delta_{ij}=m_i-m_j$ to represent the mass difference between two quarks of different flavors $i$ and $j$, the choice of $\Delta_{ij}$ depends on the studied systems, which actually implies the fact that the effective quark mass may be affected by the quark environment. Explicitly, the formula we will adopt in the present study is
\begin{eqnarray}\label{mass-4140}
M=[M_{X(4140)}-(E_{CMI})_{X(4140)}]+\sum_{ij} n_{ij}\Delta_{ij}+E_{CMI},
\end{eqnarray}
where $n_{ij}$ is an integer. In Ref. \cite{Wu:2018xdi}, the spectra of $F_{2,3,4}=|bs\bar{c}\bar{n}\rangle,\ |bn\bar{c}\bar{s}\rangle,\ |bn\bar{c}\bar{n}\rangle$ with this formula had been explored. In this article, we extend the investigation to other systems. From our  studies of various systems within the CMI model \cite{Wu:2016gas,Chen:2016ont,Wu:2017weo,Luo:2017eub,Zhou:2018pcv,Wu:2016vtq,Li:2018vhp,Wu:2018xdi}, larger tetraquark masses are usually obtained with Eq. \eqref{mass-ori} than with other methods. We treat them as upper limits of the tetraquark masses, $M_{upper}$. For comparison, we will also give their values in the present work.

\subsection{Rearrangement decays}

In each system, one may finally obtain the tetraquark wave functions and corresponding eigenvalues with the above CMI matrices. For each state, its spin$\otimes$color wave function can be written as
\begin{eqnarray}\label{Psi}
\Psi^{J=0}&=&x_1B_1+x_2B_2+x_3B_3+x_4B_4,  \nonumber\\
\Psi^{J=1}&=&x_1B_5+x_2B_6+x_3B_7+x_4B_8+c_5B_9+x_6B_{10},\nonumber\\
\Psi^{J=2}&=&x_1B_{11}+x_2B_{12}.
\end{eqnarray}
The normalization condition requires $\sum_{i=1}|x_i|^2=1$ for each tetraquark state. In Ref. \cite{Cheng:2019obk}, we used a simple model in which the Hamiltonian for rearrangement decays is just taken as a constant to investigate the decay properties of the $uudc\bar{c}$ pentaquark states. It was found that the measured ratios between decay widths of $P_c(4312)$, $P_c(4440)$, and $P_c(4457)$ can be understood well. Here, we also preliminarily investigate the decay properties of the tetraquarks within this simple model. There are two types of rearrangement decays for tetraquark states, $q_1q_2\bar{q}_3\bar{q}_4\to (q_1\bar{q}_3)_{1_c}+(q_2\bar{q}_4)_{1_c}$ and $q_1q_2\bar{q}_3\bar{q}_4\to (q_1\bar{q}_4)_{1_c}+(q_2\bar{q}_3)_{1_c}$. The final state meson-meson wave function will be recoupled to the combination of the $B_{1-12}$ bases. In doing so, we use the formulas
\begin{eqnarray}
(q_1\bar{q}_3)_{1_c}(q_2\bar{q}_4)_{1_c}&=&\frac{\sqrt6}{3}(q_1q_2)_{6_c}(\bar{q}_3\bar{q}_4)_{\bar{6}_c}-\frac{1}{\sqrt3}(q_1q_2)_{\bar{3}_c}(\bar{q}_3\bar{q}_4)_{3_c},\nonumber\\
(q_1\bar{q}_4)_{1_c}(q_2\bar{q}_3)_{1_c}&=&\frac{\sqrt6}{3}(q_1q_2)_{6_c}(\bar{q}_3\bar{q}_4)_{\bar{6}_c}+\frac{1}{\sqrt3}(q_1q_2)_{\bar{3}_c}(\bar{q}_3\bar{q}_4)_{3_c}.
\end{eqnarray}
Note that we construct the wave functions explicitly with the convention used in Refs. \cite{deSwart:1963pdg,Kaeding:1995vq} and the relative signs are different from those in Refs. \cite{Wang:2019rdo,Stancu:1991rc}. The amplitude ${\cal M}=\langle inital|H_{decay}|final\rangle=\alpha \langle inital|final\rangle$ is then obtained and the two-body decay widths can be calculated with the standard formula. Note that each system has its own constant value of $\alpha$. If we use $y_i$ ($i=1,2,...$) to denote the coefficients of tetraquark base states in the final meson-meson state, one has
\begin{eqnarray}\label{cme}
|{\cal M}|^2&=&\alpha^2[\sum_i(x_iy_i)]^2\equiv\alpha^2|\tilde{\cal M}|^2.
\end{eqnarray}
For convenience, we will call $|\tilde{\cal M}|^2$ {\it coupling matrix element} (CME) in the following discussions.

\section{Numerical results}\label{sec3}

\begin{table}[!htb]
\caption{Effective coupling parameters $C_{ij}$'s in units of MeV. The extracted effective quark masses are $m_n=361.8$ MeV, $m_s=542.4$ MeV, $m_c=1724.1$ MeV, and $m_b=5054.4$ MeV.}\label{Cij}
\centering
\begin{tabular}{p{1cm}<{\centering}p{1cm}<{\centering}p{1cm}<{\centering}p{1cm}<{\centering}p{1cm}<{\centering}|p{1cm}<{\centering}p{1cm}<{\centering}p{1cm}<{\centering}p{1cm}<{\centering}p{1cm}<{\centering}}
\thickhline\hline
$C_{ij}$    &n   &s   &c   &b   &$C_{i\bar{j}}$&$\bar{n}$&$\bar{s}$&$\bar{c}$&$\bar{b}$\\ \hline
n           &18.3&12.0&4.0 &1.3 &n             &29.9&18.7&6.6 &2.1 \\
s           &    &5.7 &4.4 &0.9 &s             &    &9.3&6.7 &2.3 \\
c           &    &    &3.2 &2.0 &c             &    &    &5.3 &3.3 \\
b           &    &    &    &1.8 &b             &    &    &    &2.9 \\ \hline\hline
\end{tabular}\\
\end{table}

\begin{table}[htbp]
\caption{Quark mass differences (units: MeV) determined with various
hadrons. Those from the extracted effective quark masses are
$\Delta_{bc}$=3330.3 MeV£¬ $\Delta_{cn}$= 1362.3 MeV, and $\Delta_{sn}$= 180.6 MeV.}\label{quarkmassdifference} \centering
\begin{tabular}{ccc|ccc|ccc}\hline\hline
Hadron&Hadron&$\Delta_{bc}$ & Hadron&Hadron&$\Delta_{cn}$& Hadron& Hadron&$\Delta_{sn}$ \\
\hline\hline
$B$ ($B^*$)&$D$ ($D^*$)&3340.2 (3340.1)&$D$ ($D^*$)&$\pi$ ($\rho$, $\omega$)&1356.4 (1357.6, 1350.2)& $K$ ($K^*$)&$\pi$ ($\rho$, $\omega$)&178.4 (178.1, 170.7)\\
$B_s$ ($B_s^*$)&$D_s$ ($D_s^*$)&3328.2 (3326.7)&$D_s$ ($D_s^*$)&$K$ ($K^*$)&1280.7 (1282.6)&          $\phi$&$K^*$&175.9\\
$B_c$&$\eta_c$ & 3259.0&                         $\eta_c$ ($J/\psi$)&$D$ ($D^*$)&1095.9 (1095.3)& $D_s$ ($D_s^*$)&$D$ ($D^*$)&102.7 (103.1)\\
$\eta_b$&$B_c$ & 3117.7&                         $B_c$&$B$&1014.6&$B_s$ ($B_s^*$)&$B$ ($B^*$)&90.6 (89.7) \\
&&  &&&&   \\
$\Lambda_b$&$\Lambda_c$, $\Sigma_c$&3333.1, 3318.6&  $\Lambda_c$&$N$&1347.5&    $\Lambda$&$N$&176.8\\
$\Sigma_b$ ($\Sigma_b^*$)&$\Sigma_c$ ($\Sigma_c^*$)&3331.1 (3329.9)&$\Sigma_c$ ($\Sigma_c^*$)&$N$ ($\Delta$)&1362.1 (1362.4)& $\Sigma$ ($\Sigma^*$)&$N$ ($\Delta$)&187.0 (186.2)\\
$\Sigma_b$&$\Lambda_c$&3345.6&   $\Xi_c^\prime$&$\Lambda$&1328.8&  $\Xi$&$\Lambda$&169.0\\
$\Xi_b$&$\Xi_c$, $\Xi_c^\prime$&3325.1, 3299.6&    $\Xi_c^\prime$ ($\Xi_c^*$)&$\Sigma$ ($\Sigma^*$)& 1318.6 (1319.8)&  $\Xi$ ($\Xi^*$)&$\Sigma$ ($\Sigma^*$)& 158.7 (159.3)\\
$\Xi_b^\prime$ ($\Xi_b^*$)&$\Xi_c^\prime$ ($\Xi_c^*$)&3323.9 (3323.2)&   $\Xi_c$&$\Lambda$, $\Sigma$&1303.3, 1293.0 &   $\Omega$&$\Xi^*$& 172.7  \\
$\Xi_b^\prime$&$\Xi_c$&3349.4&$\Omega_c$ ($\Omega_c^*$)&$\Xi$ ($\Xi^*$)&1295.9 (1273.0)&$\Xi_c^\prime$&$\Lambda_c$&158.0 \\
$\Omega_b$ &$\Omega_c$&3313.6&&&&   $\Xi_c^\prime$ ($\Xi_c^*$)&$\Sigma_c$ ($\Sigma_c^*$)&143.5 (143.5)\\
&&&&&&$\Xi_c$ &$\Lambda_c$, $\Sigma_c$&132.5, 118.0 \\
&&&&&& $\Omega_c$ &$\Xi_c$&161.6\\
&&&&&& $\Omega_c$ ($\Omega_c^*$)&$\Xi_c^\prime$ ($\Xi_c^*$)&136.0 (135.7)\\
&&&&&&$\Xi_b^\prime$&$\Lambda_b$&148.8\\
&&&&&&$\Xi_b^\prime$ ($\Xi_b^*$)&$\Sigma_b$ ($\Sigma_b^*$)&136.3 (136.8)\\
&&&&&&$\Xi_b$&$\Lambda_b$, $\Sigma_b$&124.5, 112.0\\
&&&&&&$\Omega_b$ &$\Xi_b$, $\Xi_b^\prime$&150.1, 125.7\\
&&&&&&&&\\

$\eta_b$&$\eta_c$&3188.4&     $\eta_c$ ($J/\psi$)&$\pi$  ($\rho$, $\omega$)&1226.1 (1226.4, 1222.7) &  $\phi$&$\rho$, $\omega$&177.0, 173.3\\
&&&   $\Xi_{cc}$&$N$&1287.2 &  $\Xi$ ($\Xi^*$)&$N$ ($\Delta$)&172.9 (184.3)\\
&&&&&&$\Omega$&$\Delta$, $\Sigma^*$&180.4, 177.5\\
&&&&&&$\Omega_c$&$\Lambda_c$&147.0\\
&&&&&&$\Omega_c$ ($\Omega_c^*$)&$\Sigma_c$ ($\Sigma_c^*$)&139.8, 139.6\\
&&&&&&$\Omega_b$&$\Lambda_b$, $\Sigma_b$&137.3, 121.0\\
\hline
\end{tabular}
\end{table}

The coupling parameters $C_{ij}$'s and the effective quark masses are extracted from mass splittings of ground hadrons \cite{Tanabashi:2018oca}. One may consult Refs. \cite{Wu:2018xdi,Li:2018vhp} for details about the extraction procedure and here we only list the values in Table \ref{Cij}. With such parameters and the mass of $X(4140)$, 4146.5 MeV \cite{Aaij:2016iza}, we get $M_{X(4140)}-(E_{CMI})_{X(4140)}=4231.0$ MeV by using the CMI matrix in Ref. \cite{Wu:2016gas}.

Similar to the strategy of Ref. \cite{Wu:2018xdi}, we show the extracted quark mass differences from various hadrons in Table \ref{quarkmassdifference}. The results are given in three areas. The first four rows show the values determined in the meson case where one quark is the spectator. The following fourteen rows list the values determined in the baryon case where two quarks are spectators. The last six rows correspond to the case without quark spectator or with only one quark spectator in baryons. We need to select the numerical values for the quark mass differences in the following estimations. First, we focus on $\Delta_{bc}$. From Table \ref{quarkmassdifference}, the values determined with heavy quarkonia are smaller than those with other hadrons. Notice that they will lead to smaller tetraquark masses, we choose to use $\Delta_{bc}=3340.2$ MeV. Other values around 3.3 GeV do not induce significant different conclusions in the adopted method. This feature is consistent with the fact that $\Delta_{bc}$ should not be so different for values extracted from various systems according to the heavy quark symmetry. Secondly, we concentrate on $\Delta_{cn}$. The determined values can be 1.0 GeV when a heavy quarkonium is involved. If one checks the hadron mass differences between the calculated values in the CMI model and the experimental values (see table 2 of Ref. \cite{Liu:2019zoy}), one finds that the uncertainties in heavy quarkonia are larger than those in other hadrons. If we adopt $\Delta_{cn}\approx 1.0$ GeV, the purpose to reduce the mass uncertainties cannot be achieved. We here choose $\Delta_{cn}=1280.7$ MeV which is about 80 MeV smaller than the original $m_c-m_n$ in Table \ref{Cij}. Thirdly, we choose to use $\Delta_{sn}=90.6$ MeV which results in higher tetraquark masses. No other quark mass differences are needed, which will be seen in the following discussions.

For a tetraquark state mixed with different color-spin structures, its internal two-body color-magnetic interactions are affected by the remaining quark components. We may equivalently express its mass distance from the reference mass scale with the coupling parameter $C_{ij}$ and the defined measure $K_{ij}$ \cite{Li:2018vhp,Liu:2019zoy},
\begin{eqnarray}
\label{KijCij}
\langle \widehat{H}_{CMI}\rangle&=&\sum_{i<j}X_{ij}C_{ij},\nonumber\\
E_{CMI}&=&\sum_{i<j}K_{ij}C_{ij}.
\end{eqnarray}
The value of $K_{ij}$ can be calculated when diagonalizing the CMI matrix. From the sign of $K_{ij}$, one can judge whether the effective CMI between the $i$th and $j$th quark components is attractive ($K_{ij}<0$) or repulsive ($K_{ij}>0$). From the amplitude of $K_{ij}$, one can roughly understand how important the effect would be from the change of the corresponding $C_{ij}$. Their values will be explicitly given. For convenience, we will also call these $K_{ij}$'s $K$ factors.

\begin{figure}[!h]
\centering
\begin{tabular}{ccc}
\includegraphics[width=0.33\textwidth]{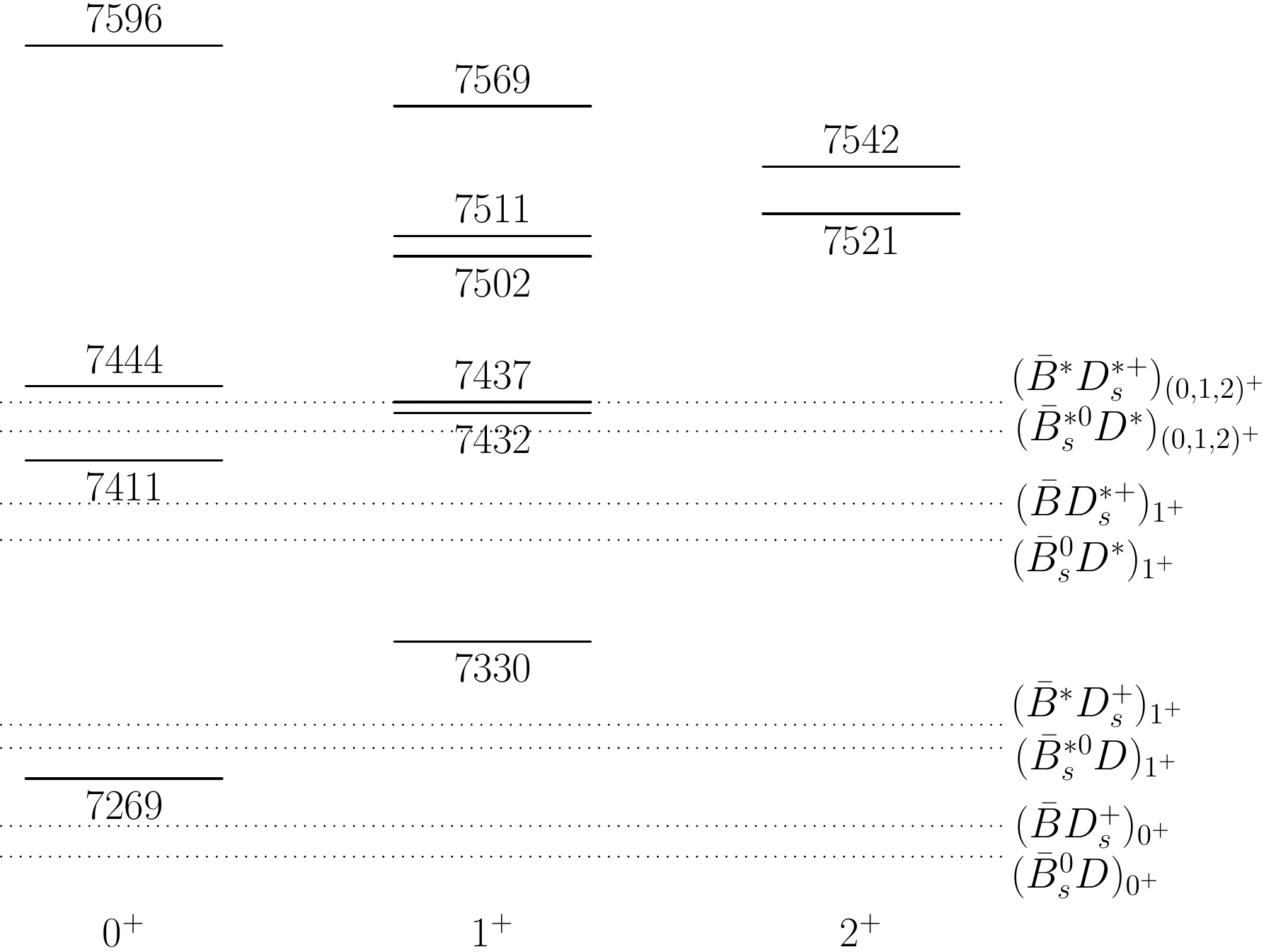}\label{ccc}&\includegraphics[width=0.33\textwidth]{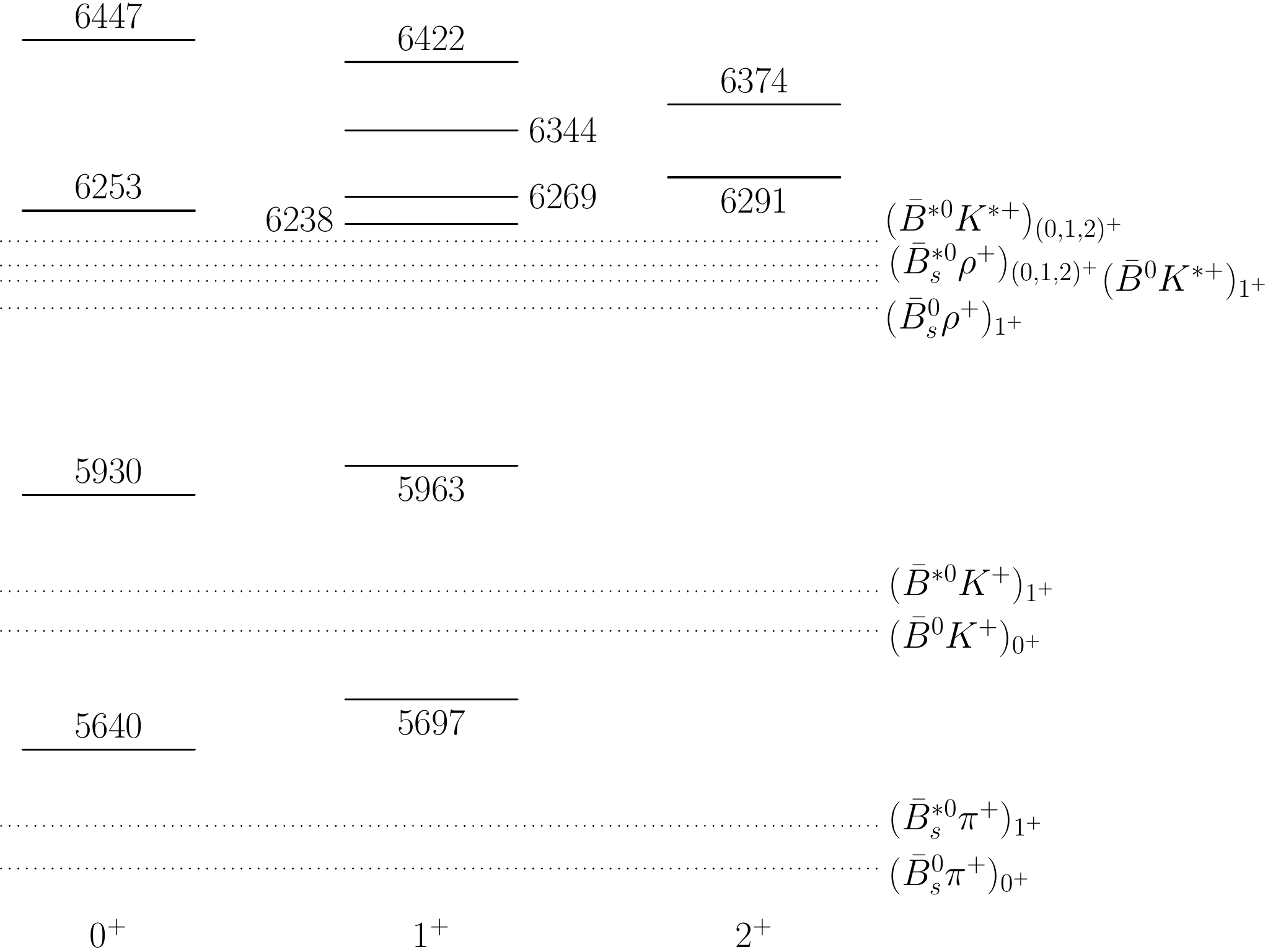}\label{ccb}&\includegraphics[width=0.33\textwidth]{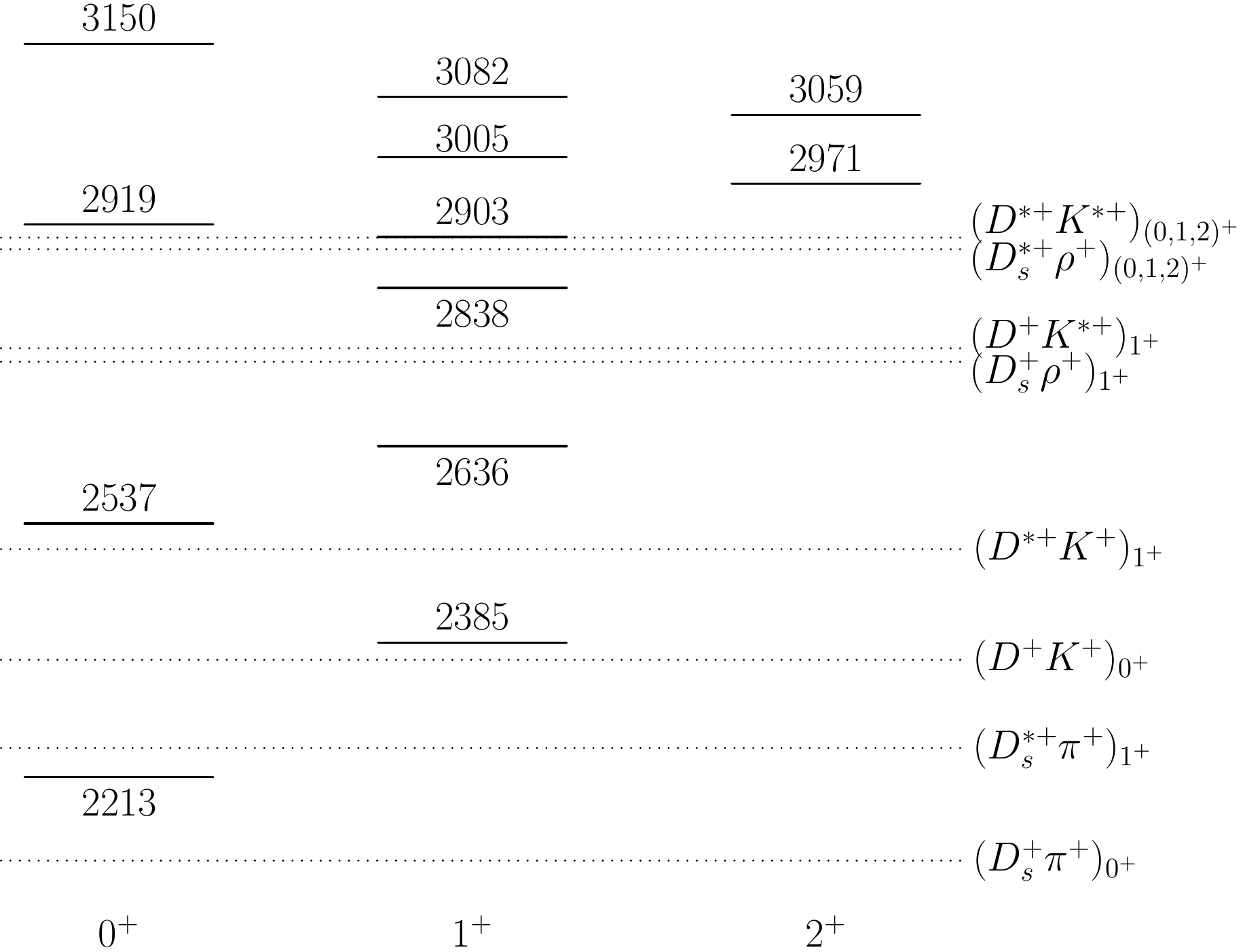}\label{ccb}\\
(a) $F_1=bc\bar{s}\bar{n}$ &(b) $F_6=bu\bar{s}\bar{d}$&(c) $F_8=cu\bar{s}\bar{d}$\\
\includegraphics[width=0.33\textwidth]{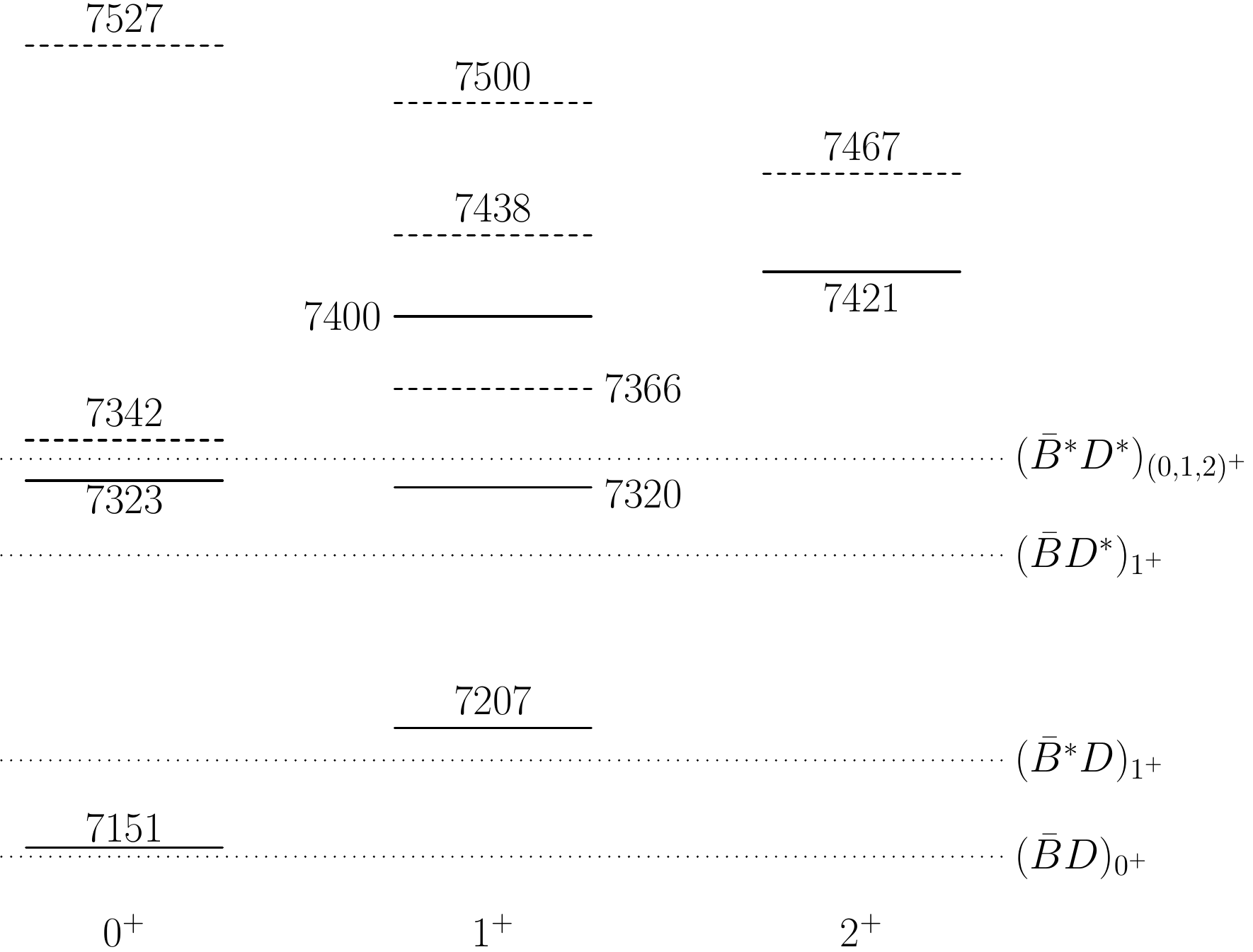}\label{ccc}&\includegraphics[width=0.33\textwidth]{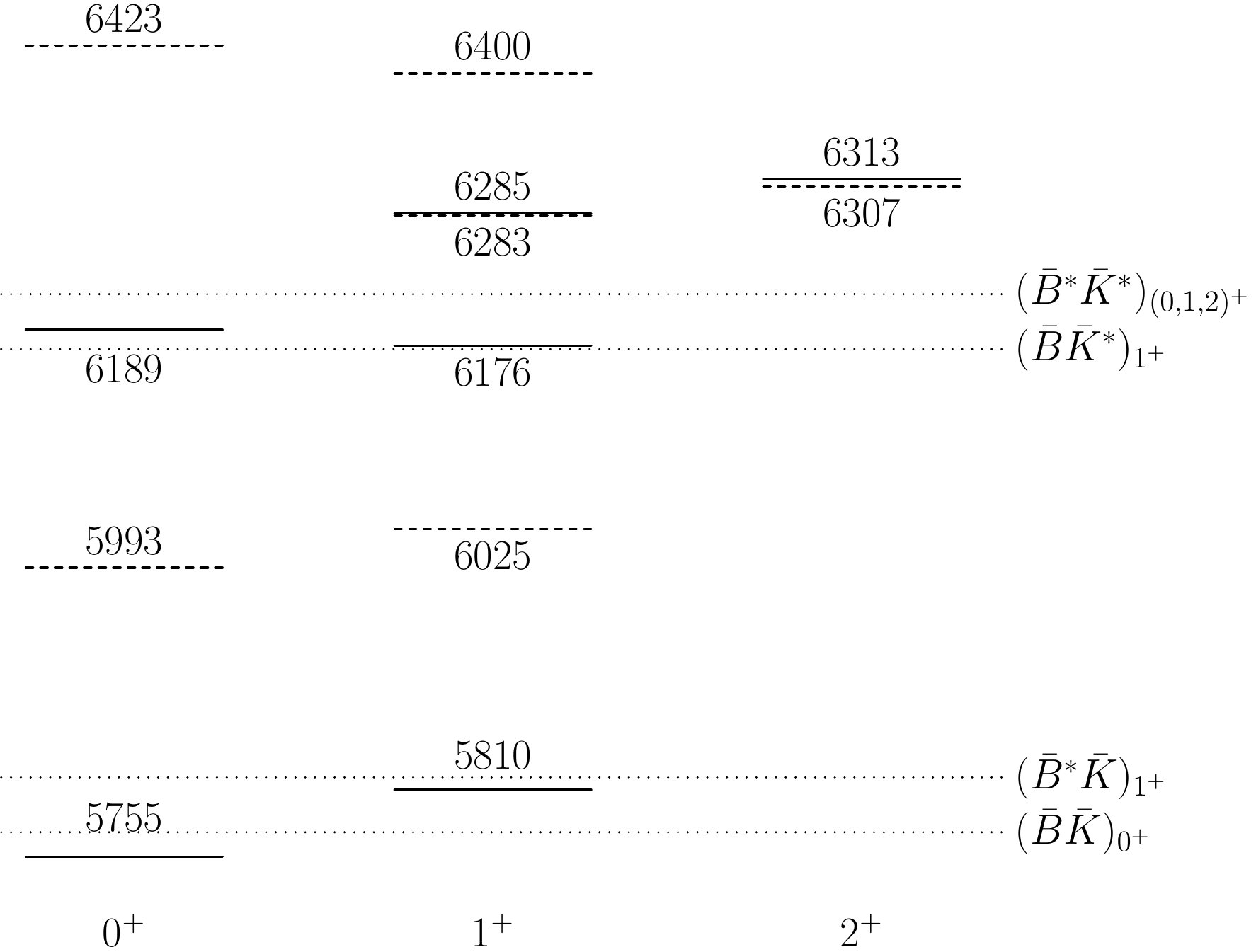}\label{ccb}&\includegraphics[width=0.33\textwidth]{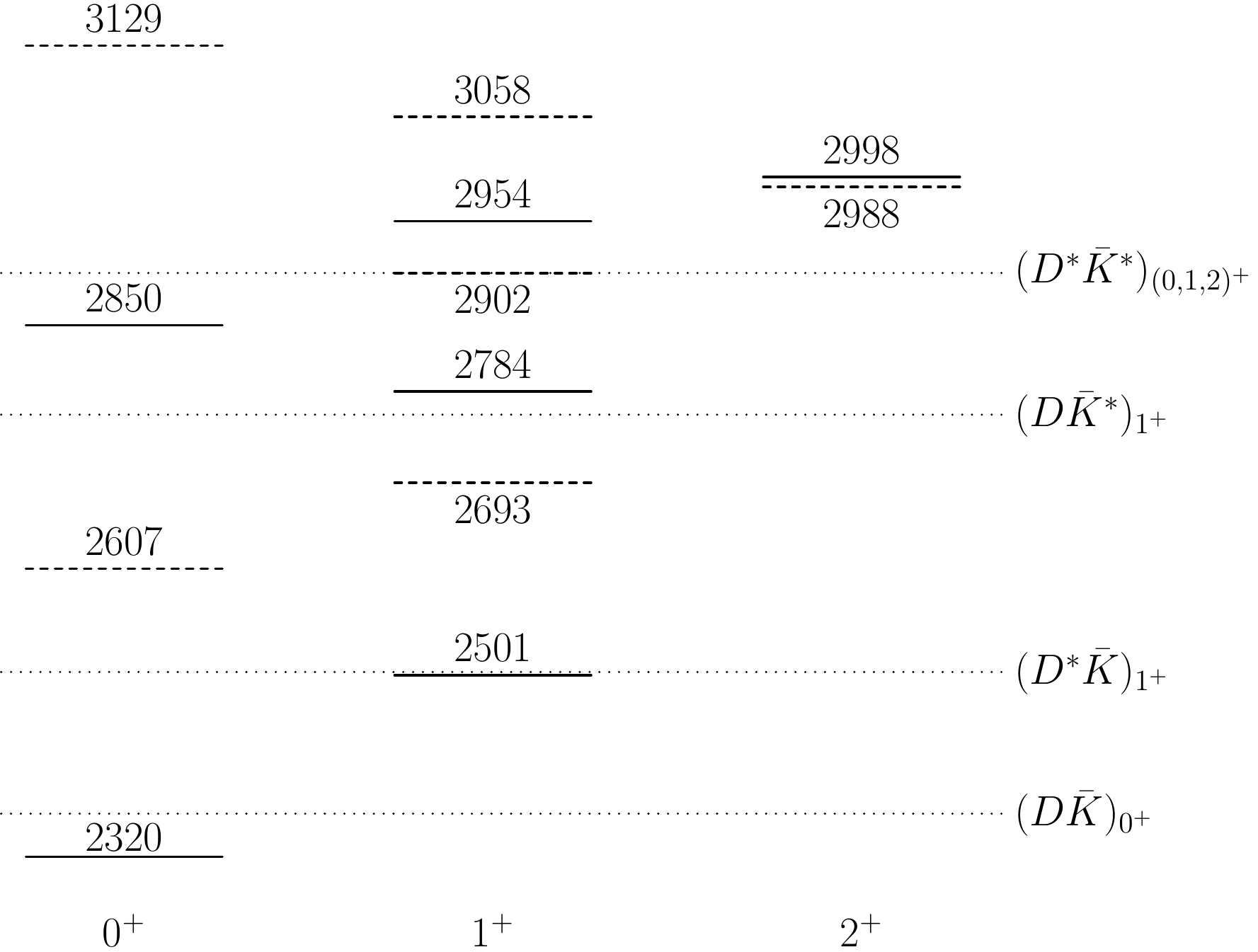}\label{ccb}\\
(d) $F_5=bc\bar{u}\bar{d}$ &(e) $F_7=bs\bar{u}\bar{d}$&(f) $F_9=cs\bar{u}\bar{d}$
\end{tabular}
\caption{Mass spectrum for the considered tetraquark states.}\label{tetraquark-fig}
\end{figure}

With the above parameters, one may derive numerical results for various systems. We collect the estimated tetraquark masses and $K_{ij}$'s for effective CMI in Tables \ref{results-mass-168} and \ref{results-mass-579}. For convenience, we also show the relative positions of the tetraquark states and related thresholds in Fig. \ref{tetraquark-fig}. In these tables and figure, we do not show the results for $F_2=bs\bar{c}\bar{n}$, $F_3=bn\bar{c}\bar{s}$, and $F_4=bu\bar{c}\bar{d}$ since they have been presented in Ref. \cite{Wu:2018xdi}. Although the systems $F_1=bc\bar{s}\bar{n}$ and $F_5=bc\bar{u}\bar{d}$ have been discussed in Ref. \cite{Luo:2017eub} and the systems $F_6=bu\bar{s}\bar{d}$, $F_7=bs\bar{u}\bar{d}$, $F_8=cu\bar{s}\bar{d}$, and $F_9=cs\bar{u}\bar{d}$ discussed in Ref. \cite{Liu:2016ogz}, the updated estimation method may result in different properties. We will discuss details about these results system by system.

The rearrangement decay widths for the considered $F_1-F_9$ systems are listed in Tables \ref{results-decay-123}, \ref{results-decay-468}, and \ref{results-decay-579}. Since $\alpha$ in the decay Hamiltonian may be different for each system, without knowing its value, one may only estimate the relative decay ratios for those states with the same $\alpha$. If we know the value of $\alpha$, all the partial decay widths for the considered states may be estimated. For this purpose, we assume that the total width equals to the sum of the partial decay widths for the rearrangement channels, i.e. $\Gamma_{total}\approx\Gamma_{sum}$, and that the parameter extracted from the LHCb decay width of $X(4140)$, $\Gamma_{X(4140)}=83^{+30}_{-25}$ MeV \cite{Aaij:2016iza}, can be applied to other tetraquark systems. In Table \ref{results-decay-cscs}, the results for the rearrangement decay widths of the $cs\bar{c}\bar{s}$ states are shown. Before the extraction of $\alpha$, we need to understand whether these results are acceptable. From Table \ref{results-decay-cscs}, $\Gamma_{X(4274)}$ is around $14.3/15.7\times83=76$ MeV when one treats it as a $cs\bar{c}\bar{s}$ tetraquark state \cite{Wu:2016gas}. This value is slightly larger than the measured $\Gamma_{X(4274)}=56^{+14}_{-16}$ MeV \cite{Aaij:2016iza}. Note that the adopted mass of $X(4274)$ (4309.4 MeV) is larger than the LHCb result (4273.3 MeV). If one adopts its measured mass, we get $\Gamma_{X(4274)}\approx 13.1/15.7*83=69$ MeV. This width is now consistent with the measured one and thus the assumptions we adopt are acceptable. However, our results in Table \ref{results-decay-cscs} cannot explain the ratio between the widths of $X(4140)$ and $X(4274)$ from Ref. \cite{Tanabashi:2018oca} where $\Gamma_{X(4274)}>\Gamma_{X(4140)}$. The experimental information about the widths is further needed to understand the nature of these two exotic states. In the present study, we use the LHCb data. With the assumption $\Gamma_{sum}=\Gamma_{total}=83$ MeV for $X(4140)$, one gets $\alpha=7.27$ GeV, a number not far from that in the hidden-charm pentaquark case \cite{Cheng:2019obk}. The following discussions about the predictions on the tetraquark widths depend on this number. If there is a $0^{++}$ $cs\bar{c}\bar{s}$ state around 4350 MeV, from Table \ref{results-decay-cscs}, its width may be comparable to that of $X(4274)$.

\begin{table}[htbp]
\caption{Numerical results for $F_1=bc\bar{s}\bar{n}$, $F_6=bu\bar{s}\bar{d}$, and $F_8=cu\bar{s}\bar{d}$ in units of MeV: CMI eigenvalues ($E_{CMI}$), upper limits for tetraquark masses ($M_{upper}$), tetraquark masses estimated with relevant thresholds and with the scale related to $X(4140)$, and $K$ factors for effective CMI. }\label{results-mass-168}
\begin{tabular}{cccccccccccc}\hline
\multicolumn{12}{c}{$F_1=bc\bar{s}\bar{n}$} \\\hline
$J^{P}$&$E_{CMI}$&$M_{upper}$&$B_sD$&$D_s\bar{B}$&$X(4140)$&$K_{bc}$&$K_{b\bar{s}}$&$K_{b\bar{n}}$&$K_{c\bar{s}}$&$K_{c\bar{n}}$&$K_{ns}$\\
$2^{+}$&61.0&7744&7438&7450&7542 &2.7&1.4&1.3&1.3&1.4&2.7\\
&40.3&7723&7417&7429&7521  &-1.3&3.3&3.4&3.4&3.3&-1.3\\
$1^{+}$&88.1&7771&7465&7477&7569  &-2.7&1.0&0.8&3.5&3.5&3.6\\
&30.6&7713&7407&7419&7511         &-1.1&-2.5&-4.6&1.9&0.6&2.6\\
&21.9&7705&7398&7411&7502         &3.4&-1.2&0.7&2.4&3.6&-2.0\\
&-43.5&7639&7333&7345&7437        &-2.7&2.1&-0.1&-4.1&-7.5&2.8\\
&-48.2&7635&7328&7340&7432        &0.2&-6.9&-4.4&0.1&3.3&-3.8\\
&-150.3&7532&7226&7238&7330       &1.5&2.7&2.8&-8.5&-8.2&-4.6\\
$0^{+}$&115.9&7799&7492&7505&7596 &3.6&3.7&3.7&3.7&3.7&3.6\\
&-36.8&7646&7340&7352&7444        &-5.1&2.1&1.9&1.9&2.1&-5.1\\
&-69.6&7613&7307&7319&7411        &3.1&-6.4&-6.4&-6.4&-6.4&3.1\\
&-212.0&7471&7165&7177&7269       &-4.2&-8.7&-8.6&-8.6&-8.7&-4.2\\
\hline\hline
\multicolumn{12}{c}{$F_6=bu\bar{s}\bar{d}$} \\\hline
$J^{P}$ & $E_{CMI}$&$M_{upper}$&$B_s\pi$&$\bar{B}K$&$X(4140)$
&$K_{bn}$&$K_{b\bar{s}}$&$K_{b\bar{n}}$&$K_{n\bar{s}}$&$K_{n\bar{n}}$&$K_{ns}$\\
$2^{+}$&174.2&6495&6194&6282&6374           &-0.6&5.2&0.8&0.8&5.2&-0.6\\
&90.9&6411&6111&6199&6291                   &1.9&-0.5&3.9&3.9&-0.5&1.9\\
$1^{+}$&222.1&6543&6242&6330&6422           &-1.9&0.9&-0.9&3.1&4.2&3.3\\
&144.6&6465&6165&6252&6344                  &2.1&-7.4&1.2&1.1&4.9&-0.8\\
&69.6&6390&6090&6178&6269                   &-2.4&-0.9&-1.7&3.1&0.3&1.0\\
&38.3&6359&6058&6146&6238                   &1.1&-2.5&-8.1&2.9&2.0&-4.7\\
&-237.0&6083&5783&5871&5963                 &-1.0&0.6&3.6&-11.2&-2.0&2.1\\
&-502.6&5818&5518&5605&5697                 &0.7&4.7&1.2&-3.8&-14.0&-2.3\\
$0^{+}$&247.5&6568&6268&6355&6447           &3.4&4.1&3.3&3.3&4.1&3.4\\
&53.0&6373&6073&6161&6253                   &-5.9&2.5&2.4&2.4&2.5&-5.9\\
&-270.3&6050&5750&5838&5930                 &2.3&-2.3&-10.9&-10.9&-2.3&2.3\\
&-560.3&5760&5460&5548&5640                 &-2.4&-13.7&-4.2&-4.2&-13.7&-2.4\\
\hline\hline
\multicolumn{12}{c}{$F_8=cu\bar{s}\bar{d}$} \\\hline
$J^{P}$ & $E_{CMI}$&$M_{upper}$&$D_s\pi$&$DK$&$X(4140)$
&$K_{cn}$&$K_{c\bar{s}}$&$K_{c\bar{n}}$&$K_{n\bar{s}}$&$K_{n\bar{n}}$&$K_{ns}$\\
$2^{+}$&198.9&3189&2891&2967&3059            &-0.7&5.1&1.0&1.0&5.1&-0.7\\
&111.3&3101&2803&2879&2971                   &2.0&-0.4&3.7&3.7&-0.4&2.0\\
$1^{+}$&222.5&3213&2914&2990&3082            &-2.7&3.0&-0.2&2.4&4.5&3.0\\
&145.1&3135&2837&2913&3005                   &3.2&-2.1&3.8&3.0&3.1&-2.3\\
&43.0&3033&2735&2811&2903                    &-1.6&-6.2&-1.0&1.2&1.9&1.4\\
&-22.0&2968&2670&2746&2838                   &0.8&-4.3&-11.4&3.4&1.8&-3.3\\
&-223.8&2766&2468&2544&2636                  &-1.7&0.1&3.1&-11.2&-1.8&2.0\\
&-475.1&2515&2217&2293&2385                  &0.5&4.8&1.0&-3.5&-14.2&-2.2\\
$0^{+}$&290.0&3280&2982&3058&3150            &3.5&4.1&3.4&3.4&4.1&3.5\\
&59.0&3049&2751&2827&2919                    &-6.0&2.6&2.4&2.4&2.6&-6.0\\
&-322.8&2667&2369&2445&2537                  &2.5&-2.7&-10.4&-10.4&-2.7&2.5\\
&-646.5&2344&2046&2121&2213                  &-2.6&-13.2&-4.8&-4.8&-13.2&-2.6\\
\hline
\end{tabular}
\end{table}

\begin{table}[htbp]
\caption{Numerical results for $F_5=bc\bar{u}\bar{d}$, $F_7=bs\bar{u}\bar{d}$, and $F_9=cs\bar{u}\bar{d}$ in units of MeV: CMI eigenvalues ($E_{CMI}$), upper limits for tetraquark masses ($M_{upper}$), tetraquark masses estimated with relevant thresholds and with the scale related to $X(4140)$, and $K$ factors for effective CMI.}\label{results-mass-579}
\begin{tabular}{ccccccccc||ccccccccc}\hline
\multicolumn{18}{c}{$F_5=bc\bar{u}\bar{d}$} \\\hline
$I(J^{P})$ & $E_{CMI}$&$M_{upper}$&$D\bar{B}$&$X(4140)$&$K_{bc}$&$K_{b\bar{n}}$&$K_{c\bar{n}}$&$K_{nn}$&$I(J^{P})$ & $E_{CMI}$&$M_{upper}$&$D\bar{B}$&$X(4140)$&$K_{bc}$&$K_{b\bar{n}}$&$K_{c\bar{n}}$&$K_{nn}$\\
$1(2^{+})$&77.4&7579&7363&7467         &2.7&2.7&2.7&2.7   &$0(2^{+})$&30.9&7533&7317&7421         &-1.3&6.7&6.7&-1.3\\
$1(1^{+})$&110.2&7612&7396&7500        &-2.5&1.5&7.0&3.6  &$0(1^{+})$&10.5&7513&7296&7400         &3.4&-2.9&6.1&-1.7\\
&47.9&7550&7334&7438                &-1.4&-6.9&2.4&2.7    &&-70.2&7432&7216&7320               &-0.0&-8.1&1.0&-3.2\\
&-24.0&7478&7262&7366               &-2.8&2.7&-12.1&3.1    &&-182.7&7319&7103&7207              &2.0&4.3&-13.7&-5.8\\
$1(0^{+})$&137.4&7639&7423&7527        &3.6&7.4&7.4&3.6   &$0(0^{+})$&-67.0&7435&7219&7323        &-4.3&2.3&2.3&-4.3\\
&-48.3&7454&7238&7342               &3.1&-12.7&-12.7&3.1    &&-238.6&7263&7047&7151              &-5.0&-15.6&-15.6&-5.0\\
\hline\hline
\multicolumn{18}{c}{$F_7=bs\bar{u}\bar{d}$} \\\hline
$I(J^P)$ & $E_{CMI}$&$M_{upper}$&$\bar{B}\bar{K}$&$X(4140)$  &$K_{bs}$&$K_{b\bar{n}}$&$K_{n\bar{s}}$&$K_{nn}$
&$I(J^P)$ & $E_{CMI}$&$M_{upper}$&$\bar{B}\bar{K}$&$X(4140)$  &$K_{bs}$&$K_{b\bar{n}}$&$K_{n\bar{s}}$&$K_{nn}$\\
$1(2^+)$&106.7&6427&6215&6307         &2.7&2.7&2.7&2.7     &$0(2^+)$&113.1&6433&6221&6313         &-1.3&6.7&6.7&-1.3\\
$1(1^+)$&200.6&6521&6308&6400         &-2.1&-0.0&7.4&3.5   &$0(1^+)$&84.9&6405&6193&6285          &2.9&-5.5&6.6&-1.6\\\
&83.5&6404&6191&6283                 &-3.2&-6.6&2.7&2.7    &&-23.7&6297&6084&6176                &1.3&-7.2&4.9&-5.5\\
&-174.7&6146&5933&6025               &-1.4&4.0&-12.8&3.1    &&-390.3&5930&5718&5810               &1.2&6.0&-18.1&-3.6\\
$1(0^+)$&223.5&6544&6331&6423         &3.5&7.5&7.5&3.5     &$0(0^+)$&-11.2&6309&6097&6189         &-5.8&4.8&4.8&-5.8\\
&-206.4&6114&5902&5993               &3.1&-12.8&-12.8&3.1    &&-445.3&5875&5663&5755               &-3.5&-18.2&-18.2&-3.5\\
\hline\hline
\multicolumn{18}{c}{$F_9=cs\bar{u}\bar{d}$} \\\hline
$I(J^P)$ & $E_{CMI}$&$M_{upper}$&$D\bar{K}$&$X(4140)$  &$K_{cs}$&$K_{c\bar{n}}$&$K_{n\bar{s}}$&$K_{nn}$
&$I(J^P)$ & $E_{CMI}$&$M_{upper}$&$D\bar{K}$&$X(4140)$  &$K_{cs}$&$K_{c\bar{n}}$&$K_{n\bar{s}}$&$K_{nn}$\\
$1(2^+)$&128.0&3118&2896&2988      &2.7&2.7&2.7&2.7     &$0(2^+)$&138.4&3128&2906&2998      &-1.3&6.7&6.7&-1.3\\
$1(1^+)$&197.9&3188&2966&3058      &-3.1&2.7&6.9&3.5    &$0(1^+)$&94.1&3084&2862&2954       &3.5&2.7&6.0&-2.8\\
&42.4&3033&2810&2902              &-0.9&-8.0&2.7&2.7    &&-75.4&2915&2692&2784             &0.7&-15.4&5.5&-4.3\\
&-166.3&2824&2601&2693            &-2.6&2.7&-12.3&3.1    &&-359.1&2631&2409&2501            &1.1&6.0&-18.1&-3.6\\
$1(0^+)$&269.5&3260&3037&3129      &3.5&7.5&7.5&3.5     &$0(0^+)$&-9.8&2980&2758&2850       &-5.8&4.8&4.8&-5.8\\
&-253.1&2737&2515&2607            &3.1&-12.8&-12.8&3.1    &&-539.4&2451&2228&2320            &-3.5&-18.2&-18.2&-3.5\\
\hline
\end{tabular}
\end{table}

\begin{table}[htbp]
\caption{Rearrangement decays for $F_1=bc\bar{s}\bar{n}$, $F_2=bs\bar{c}\bar{n}$, and $F_3=bn\bar{c}\bar{s}$. The numbers in the parentheses are ($100 |{\cal M}|^2/\alpha^2$, $10^7\Gamma/\alpha^2$). The symbol ``$-$'' means that the decay is forbidden. Decay channels from left to right are presented with increasing thresholds.}\label{results-decay-123}
\begin{tabular}{cccccccc}\hline
$F_1$ states&\multicolumn{6}{c}{Channels}&$\Gamma_{sum}$\\
$J^P=2^+$&$\bar{B}_s^{*0}D^*$&$\bar{B}^*D_s^{*+}$&&&&&\\
7542&(34.6,1.4)&(32.1,1.3)&&&&&2.7\\
7521&(65.4,2.5)&(67.9,2.4)&&&&&4.9\\
$J^P=1^+$&$\bar{B}_s^{*0}D$&$\bar{B}^*D_s^+$&$\bar{B}_s^0D^*$&$\bar{B}D_s^{*+}$&$\bar{B}_s^{*0}D^*$&$\bar{B}^*D_s^{*+}$&\\
7569&(0.6,0.0)&(0.6,0.0)&(8.8,0.5)&(9.6,0.5)&(46.7,2.1)&(45.1,2.0)&5.2\\
7511&(1.6,0.1)&(1.1,0.1)&(22.2,1.0)&(35.0,1.5)&(1.3,0.0)&(6.9,0.2)&2.9\\
7502&(1.2,0.1)&(1.9,0.1)&(24.6,1.1)&(13.2,0.5)&(44.4,1.5)&(40.4,1.3)&4.6\\
7437&(45.9,2.2)&(31.1,1.5)&(0.4,0.0)&(9.9,0.3)&(0.0,0.0)&(1.5,0.0)&3.9\\
7432&(1.6,0.1)&(14.2,0.7)&(44.0,1.3)&(32.1,0.8)&(7.4,0.1)&(5.9,$-$)&2.9\\
7330&(49.2,1.3)&(51.1,1.2)&(0.1,$-$)&(0.1,$-$)&(0.2,$-$)&(0.2,$-$)&2.6\\
$J^P=0^+$&$\bar{B}_s^0D$&$\bar{B}D_s^+$&$\bar{B}_s^{*0}D^*$&$\bar{B}^*D_s^{*+}$&&&\\
7596&(0.3,0.0)&(0.3,0.0)&(55.8,2.8)&(55.5,2.7)&&&5.5\\
7444&(5.7,0.3)&(6.4,0.3)&(41.7,0.7)&(41.5,0.4)&&&1.8\\
7411&(41.3,2.1)&(41.5,2.1)&(2.4,$-$)&(2.9,$-$)&&&4.2\\
7269&(52.8,1.2)&(51.8,1.0)&(0.0,$-$)&(0.0,$-$)&&&2.2\\
\hline
$F_2$ states&\multicolumn{6}{c}{Channels}&$\Gamma_{sum}$\\
$J^P=2^+$&$B_c^{*-}\bar{K}^*$&$\bar{B}^*D_s^{*-}$&&&&&\\
7599&(99.3,5.5)&(17.1,0.8)&&&&&6.3\\
7513&(0.7,0.0)&(82.9,2.8)&&&&&2.9\\
$J^P=1^+$&$B_c^{*-}\bar{K}$&$B_c^-\bar{K}^*$&$B_c^{*-}\bar{K}^*$&$\bar{B}^*D_s^-$&$\bar{B}D_s^{*-}$&$\bar{B}^*D_s^{*-}$&\\
7604&(0.0,0.0)&(0.8,0.1)&(91.9,5.1)&(1.6,0.1)&(9.3,0.5)&(5.3,0.3)&6.1\\
7552&(0.1,0.0)&(44.7,2.6)&(6.4,0.3)&(0.8,0.1)&(0.0,0.0)&(60.7,2.5)&5.5\\
7510&(0.2,0.0)&(49.3,2.7)&(0.8,0.0)&(1.9,0.1)&(0.2,0.0)&(25.7,0.9)&3.7\\
7482&(0.0,0.0)&(4.5,0.2)&(0.5,0.0)&(3.9,0.2)&(83.6,3.1)&(3.3,0.1)&3.7\\
7392&(2.2,0.1)&(0.7,0.0)&(0.5,0.0)&(82.7,3.2)&(4.6,0.0)&(2.0,$-$)&3.4\\
7194&(97.6,4.9)&(0.0,0.0)&(0.0,$-$)&(9.0,$-$)&(2.3,$-$)&(3.0,$-$)&4.9\\
$J^P=0^+$&$B_c^-\bar{K}$&$B_c^{*-}\bar{K}^*$&$\bar{B}D_s^-$&$\bar{B}^*D_s^{*-}$&&&\\
7620&(0.0,0.0)&(79.1,4.5)&(0.8,0.1)&(30.0,1.5)&&&6.1\\
7527&(0.2,0.0)&(19.5,1.0)&(4.9,0.3)&(60.7,2.2)&&&3.5\\
7355&(3.3,0.2)&(1.4,0.0)&(83.5,3.4)&(4.6,$-$)&&&3.7\\
7120&(96.6,4.9)&(0.0,$-$)&(10.8,$-$)&(4.8,$-$)&&&4.9\\
\hline
$F_3$ states&\multicolumn{6}{c}{Channels}&$\Gamma_{sum}$\\
$J^P=2^+$&$B_c^{*-}K^*$&$\bar{B}_s^{*0}\bar{D}^*$&&&&&\\
7598&(99.5,5.5)&(16.0,0.8)&&&&&6.3\\
7515&(0.5,0.0)&(84.0,3.1)&&&&&3.1\\
$J^P=1^+$&$B_c^{*-}K$&$B_c^-K^*$&$B_c^{*-}K^*$&$\bar{B}_s^{*0}\bar{D}$&$\bar{B}_s^0\bar{D}^*$&$\bar{B}_s^{*0}\bar{D}^*$&\\
7603&(0.0,0.0)&(0.9,0.1)&(91.9,5.1)&(1.8,0.1)&(9.3,0.5)&(5.3,0.3)&6.1\\
7553&(0.1,0.0)&(45.5,2.6)&(6.5,0.3)&(0.9,0.1)&(0.0,0.0)&(59.6,2.6)&5.6\\
7509&(0.2,0.0)&(47.4,2.6)&(0.7,0.0)&(1.8,0.1)&(0.0,0.0)&(27.2,1.0)&3.7\\
7479&(0.1,0.0)&(5.6,0.3)&(0.5,0.0)&(4.9,0.3)&(82.7,3.3)&(3.0,0.1)&3.9\\
7394&(2.1,0.1)&(0.7,0.0)&(0.4,0.0)&(81.6,3.3)&(5.5,0.1)&(1.9,$-$)&3.6\\
7194&(97.6,4.9)&(0.0,0.0)&(0.0,$-$)&(9.0,$-$)&(2.4,$-$)&(2.9,$-$)&4.9\\
$J^P=0^+$&$B_c^-K$&$\bar{B}_s^0\bar{D}$&$B_c^{*-}K^*$&$\bar{B}_s^{*0}\bar{D}^*$&&&\\
7622&(0.0,0.0)&(0.8,0.1)&(77.5,4.4)&(31.8,1.7)&&&6.2\\
7523&(0.2,0.0)&(5.4,0.3)&(21.3,1.1)&(58.7,2.2)&&&3.7\\
7354&(3.7,0.2)&(82.3,3.5)&(1.3,0.0)&(4.9,$-$)&&&3.8\\
7119&(96.1,4.8)&(11.5,$-$)&(0.0,$-$)&(4.6,$-$)&&&4.8\\
\hline
\end{tabular}
\end{table}

\begin{table}[htbp]
\caption{Rearrangement decays for isovector $F_4=bu\bar{c}\bar{d}$, $F_6=bu\bar{s}\bar{d}$, and $F_8=cu\bar{s}\bar{d}$. The numbers in the parentheses are ($100 |{\cal M}|^2/\alpha^2$, $10^7\Gamma/\alpha^2\cdot$MeV). The symbol ``$-$'' means that the decay is forbidden. Decay channels from left to right are presented with increasing thresholds.}\label{results-decay-468}
\begin{tabular}{cccccccc}\hline
$F_4$ states&\multicolumn{6}{c}{Channels}&$\Gamma_{sum}$\\
$J^P=2^+$&$B_c^{*-}\rho^+$&$\bar{B}^{*0}\bar{D}^{*0}$&&&&&\\
7567&(99.8,6.0)&(13.9,0.8)&&&&&6.8\\
7415&(0.2,0.0)&(86.1,3.1)&&&&&3.1\\
$J^P=1^+$&$B_c^{*-}\pi^+$&$B_c^-\rho^+$&$B_c^{*-}\rho^+$&$\bar{B}^{*0}\bar{D}^0$&$\bar{B}^0\bar{D}^{*0}$&$\bar{B}^{*0}\bar{D}^{*0}$&\\
7570&(0.0,0.0)&(0.4,0.0)&(96.8,5.9)&(2.5,0.2)&(8.4,0.5)&(1.7,0.1)&6.7\\
7509&(0.0,0.0)&(80.3,5.0)&(1.6,0.1)&(0.0,0.0)&(0.1,0.0)&(31.1,1.6)&6.7\\
7448&(0.0,0.0)&(18.9,1.1)&(1.3,0.1)&(6.9,0.4)&(10.4,0.5)&(46.7,2.0)&4.1\\
7388&(0.1,0.0)&(0.2,0.0)&(0.1,0.0)&(3.2,0.2)&(72.1,2.9)&(14.0,0.4)&3.5\\
7296&(0.6,0.0)&(0.3,0.0)&(0.2,0.0)&(81.8,3.3)&(6.4,0.1)&(2.6,$-$)&3.4\\
6926&(99.3,4.5)&(0.0,$-$)&(0.0,$-$)&(5.6,$-$)&(2.6,$-$)&(3.9,$-$)&4.5\\
$J^P=0^+$&$B_c^-\pi^+$&$B_c^{*-}\rho^+$&$\bar{B}^0\bar{D}^0$&$\bar{B}^{*0}\bar{D}^{*0}$&&&\\
7584&(0.0,0.0)&(87.8,5.4)&(1.6,0.1)&(20.6,1.2)&&&6.7\\
7459&(0.1,0.0)&(11.5,0.6)&(7.3,0.5)&(67.3,2.9)&&&4.1\\
7258&(1.2,0.1)&(0.7,0.0)&(84.2,3.6)&(6.2,$-$)&&&3.7\\
6854&(98.8,4.5)&(0.0,$-$)&(6.8,$-$)&(5.9,$-$)&&&4.5\\
\hline
$F_6$ states&\multicolumn{6}{c}{Channels}&$\Gamma_{sum}$\\
$J^P=2^+$&$\bar{B}_s^{*0}\rho^+$&$\bar{B}^{*0}K^{*+}$&&&&&\\
6374&(97.0,4.9)&(24.1,1.2)&&&&&6.1\\
6291&(3.0,0.1)&(75.9,2.6)&&&&&2.7\\
$J^P=1^+$&$\bar{B}_s^{*0}\pi^+$&$\bar{B}^{*0}K^+$&$\bar{B}_s^0\rho^+$&$\bar{B}^0K^{*+}$&$\bar{B}_s^{*0}\rho^+$&$\bar{B}^{*0}K^{*+}$&\\
6422&(0.0,0.0)&(0.4,0.0)&(12.1,0.8)&(13.6,0.8)&(55.9,3.2)&(27.7,1.5)&6.4\\
6344&(0.0,0.0)&(0.2,0.0)&(57.2,3.1)&(6.4,0.3)&(32.0,1.5)&(23.9,1.1)&6.0\\
6269&(0.2,0.0)&(0.3,0.0)&(13.5,0.6)&(23.6,0.9)&(0.6,0.0)&(37.5,1.1)&2.6\\
6238&(0.3,0.0)&(3.1,0.2)&(16.2,0.6)&(53.4,1.7)&(10.2,0.3)&(6.9,0.1)&3.0\\
5963&(12.4,0.7)&(72.4,3.1)&(1.0,$-$)&(2.0,$-$)&(1.2,$-$)&(2.4,$-$)&3.8\\
5697&(87.0,2.6)&(23.7,$-$)&(0.0,$-$)&(1.1,$-$)&(0.0,$-$)&(1.6,$-$)&2.6\\
$J^P=0^+$&$\bar{B}_s^0\pi^+$&$\bar{B}^0K^+$&$\bar{B}_s^{*0}\rho^+$&$\bar{B}^{*0}K^{*+}$&&&\\
6447&(0.0,0.0)&(0.3,0.0)&(65.1,3.9)&(44.6,2.6)&&&6.6\\
6253&(0.6,0.1)&(3.4,0.3)&(32.6,1.0)&(48.8,1.1)&&&2.4\\
5930&(14.1,0.8)&(70.5,3.2)&(2.3,$-$)&(4.2,$-$)&&&4.0\\
5640&(85.2,2.4)&(25.8,$-$)&(0.1,$-$)&(2.5,$-$)&&&2.4\\
\hline
$F_8$ states&\multicolumn{6}{c}{Channels}&$\Gamma_{sum}$\\
$J^P=2^+$&$D_s^{*+}\rho^+$&$D^{*+}K^{*+}$&&&&&\\
3059&(95.8,18.5)&(27.0,5.2)&&&&&23.6\\
2971&(4.2,0.6)&(73.0,9.7)&&&&&10.3\\
$J^P=1^+$&$D_s^{*+}\pi^+$&$D^{*+}K^+$&$D_s^+\rho^+$&$D^+K^{*+}$&$D_s^{*+}\rho^+$&$D^{*+}K^{*+}$&\\
3082&(0.0,0.0)&(0.8,0.3)&(4.5,1.2)&(9.3,2.5)&(71.4,14.5)&(22.2,4.5)&23.0\\
3005&(0.3,0.1)&(1.3,0.4)&(22.8,5.6)&(1.3,0.3)&(26.5,4.3)&(62.1,9.9)&20.7\\
2903&(0.7,0.2)&(0.0,0.0)&(44.6,9.1)&(12.0,2.4)&(0.0,0.0)&(12.7,0.1)&11.8\\
2838&(0.0,0.0)&(3.2,0.9)&(26.4,4.3)&(71.3,10.9)&(1.3,$-$)&(0.1,$-$)&16.0\\
2636&(10.5,2.7)&(72.9,14.1)&(1.8,$-$)&(4.4,$-$)&(0.8,$-$)&(1.5,$-$)&16.8\\
2385&(88.4,13.7)&(21.8,$-$)&(0.0,$-$)&(1.7,$-$)&(0.0,$-$)&(1.4,$-$)&13.7\\
$J^P=0^+$&$D_s^+\pi^+$&$D^+K^+$&$D_s^{*+}\rho^+$&$D^{*+}K^{*+}$&&&\\
3150&(0.0,0.0)&(0.2,0.1)&(63.4,14.5)&(46.2,10.6)&&&25.2\\
2919&(0.7,0.3)&(3.1,1.1)&(34.0,3.0)&(47.5,3.2)&&&7.5\\
2537&(17.1,5.2)&(67.2,16.1)&(2.5,$-$)&(4.1,$-$)&&&21.3\\
2213&(82.1,12.7)&(29.4,$-$)&(0.1,$-$)&(2.1,$-$)&&&12.7\\
\hline
\end{tabular}
\end{table}

\begin{table}[htbp]
\caption{Rearrangement decays for $F_5=bc\bar{u}\bar{d}$, $F_7=bs\bar{u}\bar{d}$, and $F_9=cs\bar{u}\bar{d}$. The numbers in the parentheses are ($100 |{\cal M}|^2/\alpha^2$, $10^7\Gamma/\alpha^2\cdot$MeV). The symbol ``$-$'' means that the decay is forbidden. Note that the widths for $\bar{B}^{(*)}D^{(*)}$ include results for both $B^{(*)-}D^{(*)+}$ and $\bar{B}^{(*)0}D^{(*)0}$ channels, the widths for $\bar{B}^{(*)}\bar{K}^{(*)}$ include results for both $B^{(*)-}\bar{K}^{(*)0}$ and  $\bar{B}^{(*)0}K^{(*)-}$ channels, and the widths for $D^{(*)}\bar{K}^{(*)}$ include results for both $D^{(*)0}\bar{K}^{(*)0}$ and $D^{(*)+}K^{(*)-}$ channels. Decay channels from left to right are presented with increasing thresholds.}\label{results-decay-579}
\begin{tabular}{ccccc||ccccc}\hline
$F_5$ states&\multicolumn{3}{c}{Channels}&$\Gamma_{sum}$   &$F_5$ states&\multicolumn{3}{c}{Channels}&$\Gamma_{sum}$\\
$I(J^P)=1(2^+)$&$\bar{B}^*D^*$&&&                           &$I(J^P)=0(2^+)$&$\bar{B}^*D^*$&&&\\
7467&(33.3,3.0)&&&3.0                                     &7421&(66.7,4.9)&&&4.9\\
$I(J^P)=1(1^+)$&$\bar{B}^*D$&$\bar{B}D^*$&$\bar{B}^*D^*$&   &$I(J^P)=0(1^+)$&$\bar{B}^*D$&$\bar{B}D^*$&$\bar{B}^*D^*$&\\
7500&(0.8,0.1)&(10.3,1.2)&(45.6,4.6)&5.8                  &7401&(1.5,0.2)&(24.9,2.1)&(38.6,2.5)&4.8\\
7438&(1.2,0.1)&(28.2,2.7)&(4.1,0.3)&3.2                   &7320&(15.3,1.4)&(32.1,1.4)&(9.8,$-$)&2.8\\
7366&(39.6,4.1)&(3.2,0.2)&(0.3,0.0)&4.3                   &7207&(41.5,1.3)&(1.4,$-$)&(1.6,$-$)&1.3\\
$I(J^P)=1(0^+)$&$\bar{B}D$&$\bar{B}^*D^*$&&                 &$I(J^P)=0(0^+)$&$\bar{B}D$&$\bar{B}^*D^*$&&\\
7527&(0.4,0.1)&(56.2,6.0)&&6.1                            &7323&(11.2,1.2)&(40.6,$-$)&&1.2\\
7342&(41.2,4.5)&(2.2,0.1)&&4.6                            &7151&(47.1,0.8)&(1.0,$-$)&&0.8\\
\hline
$F_7$ states&\multicolumn{3}{c}{Channels}&$\Gamma_{sum}$                        &$F_7$ states&\multicolumn{3}{c}{Channels}&$\Gamma_{sum}$\\
$I(J^P)=1(2^+)$&$\bar{B}^*\bar{K}^*$&&&                                         &$I(J^P)=0(2^+)$&$\bar{B}^*\bar{K}^*$&&&\\
6307&(33.3,2.5)&&&2.5                                                         &6313&(66.7,5.2)&&&5.2\\
$I(J^P)=1(1^+)$&$\bar{B}^*\bar{K}$&$\bar{B}\bar{K}^*$&$\bar{B}^*\bar{K}^*$&&$I(J^P)=0(1^+)$&$\bar{B}^*\bar{K}$&$\bar{B}\bar{K}^*$&$\bar{B}^*\bar{K}^*$&\\
6400&(0.2,0.0)&(12.9,1.5)&(42.0,4.5)&6.0                                      &6285&(0.2,0.0)&(31.4,2.7)&(33.8,2.2)&4.9\\
6283&(0.0,0.0)&(27.3,2.3)&(6.3,0.4)&2.7                                       &6176&(2.9,0.4)&(26.8,0.4)&(16.1,$-$)&0.8\\
6025&(41.4,4.2)&(1.5,$-$)&(1.7,$-$)&4.2                                       &5810&(55.2,$-$)&(0.1,$-$)&(0.1,$-$)&0.0\\
$I(J^P)=1(0^+)$&$\bar{B}\bar{K}$&$\bar{B}^*\bar{K}^*$&&                         &$I(J^P)=0(0^+)$&$\bar{B}\bar{K}$&$\bar{B}^*\bar{K}^*$&&\\
6423&(0.2,0.0)&(55.1,6.2)&&6.2                                                &6189&(2.8,0.4)&(41.4,$-$)&&0.4\\
5994&(41.5,4.4)&(3.2,$-$)&&4.4                                                &5755&(55.5,$-$)&(0.2,$-$)&&0.0\\
\hline
$F_9$ states&\multicolumn{3}{c}{Channels}&$\Gamma_{sum}$&$F_9$ states&\multicolumn{3}{c}{Channels}&$\Gamma_{sum}$\\
$I(J^P)=1(2^+)$&$D^*\bar{K}^*$&&&                             &$I(J^P)=0(2^+)$&$D^*\bar{K}^*$&&&\\
2988&(33.3,9.8)&&&9.8                                       &2998&(66.7,20.6)&&&20.6\\
$I(J^P)=1(1^+)$&$D^*\bar{K}$&$D\bar{K}^*$&$D^*\bar{K}^*$&     &$I(J^P)=0(1^+)$&$D^*\bar{K}$&$D\bar{K}^*$&$D^*\bar{K}^*$&\\
3058&(0.4,0.2)&(7.1,3.8)&(46.8,17.9)&21.9                   &2954&(1.3,0.8)&(10.6,4.8)&(47.6,11.0)&16.6\\
2902&(1.1,0.7)&(30.6,12.2)&(2.3,$-$)&12.8                   &2784&(1.8,1.0)&(47.6,8.2)&(2.4,$-$)&9.2\\
2693&(40.2,18.1)&(4.0,$-$)&(0.9,$-$)&18.1                   &2501&(55.2,$-$)&(0.1,$-$)&(0.1,$-$)&0.0\\
$I(J^P)=1(0^+)$&$D\bar{K}$&$D^*\bar{K}^*$&&                   &$I(J^P)=0(0^+)$&$D\bar{K}$&$D^*\bar{K}^*$&&\\
3129&(0.2,0.1)&(55.1,24.5)&&24.6                            &2850&(2.7,1.9)&(41.4,$-$)&&1.9\\
2607&(41.5,22.8)&(3.2,$-$)&&22.8                            &2320&(55.6,$-$)&(0.3,$-$)&&0.0\\
\hline
\end{tabular}
\end{table}

\begin{table}[htbp]
\caption{Rearrangement decays for $cs\bar{c}\bar{s}$ states. The numbers in the parentheses are ($100 |{\cal M}|^2/\alpha^2$, $10^7\Gamma/\alpha^2\cdot$MeV). The symbol ``$-$'' means that the decay is forbidden. Decay channels from left to right are presented with increasing thresholds.}\label{results-decay-cscs}
\begin{tabular}{c|ccccc}\hline\hline
States&\multicolumn{4}{c}{Channels}&$\Gamma_{sum}$\\\hline
$J^{PC}=2^{++}$&$J/\psi\phi$&$D_s^{*+}D_s^{*-}$&&&\\
4313.2&(80.6,9.7)&(51.1,4.8)&&&14.5\\
4291.2&(19.4,2.2)&(48.9,4.0)&&&6.2\\
$J^{PC}=1^{++}$&$D_sD_s^*$&$J/\psi\phi$&&&\\
4309.4&(8.0,2.4)&(99.7,11.9)&&&14.3\\
4146.5&(92.0,15.7)&(0.3,0.0)&&&15.7\\
$J^{PC}=1^{+-}$&$\eta_c\phi$&$D_sD_s^*$&$D_s^{*+}D_s^{*-}$&&\\
4304.8&(7.8,1.2)&(0.2,0.1)&(98.1,8.7)&&10.0\\
4222.3&(35.4,4.7)&(24.6,5.9)&(1.2,$-$)&&10.6\\
4164.7&(56.6,6.6)&(46.8,8.9)&(0.1,$-$)&&15.5\\
4095.7&(0.3,0.0)&(28.4,2.4)&(0.6,$-$)&&2.4\\
$J^{PC}=0^{++}$&$D_s^+D_s^-$&$J/\psi\phi$&$D_s^{*+}D_s^{*-}$&&\\
4366.9&(0.1,0.0)&(56.7,7.6)&(53.1,6.1)&&13.7\\
4248.0&(2.2,0.4)&(39.7,4.0)&(42.4,2.1)&&6.5\\
4079.7&(48.0,6.1)&(3.2,$-$)&(3.7,$-$)&&6.1\\
3944.4&(49.7,1.6)&(0.4,$-$)&(0.8,$-$)&&1.6\\
\hline
\end{tabular}
\end{table}

\subsection{Systems composed of $b$, $c$, $s$, and $n$ quarks}

In this case, we have three systems, $F_1=bc\bar{s}\bar{n}$, $F_2=|bs\bar{c}\bar{n}\rangle$, and $F_3=|bn\bar{c}\bar{s}\rangle$. The mass formula for them reads
\begin{eqnarray}
M=[M_{X(4140)}-(E_{CMI})_{X(4140)}]+\Delta_{bc}-\Delta_{sn}+E_{CMI}.
\end{eqnarray}
The $F_1$ states are partners of the predicted $T_{cc}=cc\bar{u}\bar{d}$ tetraquark. They were considered previously in Ref. \cite{Luo:2017eub} with the same CMI model. The $F_2$ and $F_3$ states considered in Ref. \cite{Wu:2018xdi} are partners of the $cs\bar{c}\bar{s}$ tetraquarks. Now, with the present mass formula, the mass splittings between all these involved states are determined by the color-spin interactions. Once an exotic tetraquark containing $b$, $c$, $s$, and $n$ quarks could be observed, one may use such a state to correct the tetraquark masses obtained here.

Compared with the $F_1$ results using the $B_sD$ threshold (adopted in Ref. \cite{Luo:2017eub}), the updated masses are 100 MeV higher, but are still 200 MeV lower than the theoretical upper limits. If the present results are more realistic, the lowest $0^+$ and $1^+$ states without strong decays in Ref. \cite{Luo:2017eub} are not stable any longer. Here ``realistic'' means that the masses are closer to the measured ones by future experiments. From the values of $K_{ij}$'s, the mass splittings are not affected significantly by the uncertainties of the coupling parameters. We have argued in Ref. \cite{Wu:2018xdi} that states with negative $K_{qq}$ and positive $K_{q\bar{q}}$ are probably relatively stable. One will see that parts of them satisfy this feature.

From table \ref{results-decay-123}, the $F_1$, $F_2$, and $F_3$ states have comparable widths. If we adopt $\alpha=7.27$ MeV, the widths of these states are about $10\sim 33$ MeV. If such tetraquarks do exist, one may conclude that all of them are measurable according to the present model calculation. We now examine the decay properties of states in each system.

\subsubsection{$F_1=bc\bar{s}\bar{n}$ states}

The lowest $bc\bar{s}\bar{n}$ tetraquark is a scalar state and it has two rearrangement decay channels $\bar{B}_s^0D$ and $\bar{B}D_s^+$ with almost equal coupling matrix elements and partial widths. The second lowest $0^+$ $bc\bar{s}\bar{n}$ shares the same channels with the lowest one. Although the coupling matrix elements for decay are smaller than those of the lowest tetraquark, the larger phase spaces lead to larger partial widths. The second highest $0^+$ $bc\bar{s}\bar{n}$ has four rearrangement channels with comparable partial widths. The couplings with the channels $\bar{B}_s^0D$ and $\bar{B}D_s^+$ are weak but the large phase spaces ensure nonvanishing partial widths. Although the phase spaces for $\bar{B}_s^{*0}D^*$ and $\bar{B}^*D_s^{*+}$ are small, the large coupling matrix elements for decay result in similar widths to the channels $\bar{B}_s^0D$ and $\bar{B}D_s^+$. For the highest $0^+$ $bc\bar{s}\bar{n}$, the stronger couplings with $\bar{B}_s^{*0}D^*$ and $\bar{B}^*D_s^{*+}$ and the larger phase spaces for these two channels give the largest width. In Ref. \cite{Wu:2018xdi}, we have found that the second highest $0^+$ $Qq\bar{Q}\bar{q}$ tetraquark usually has a relatively stable structure based on the signs of $K_{ij}$'s. Now, the second highest state of these four $bc\bar{s}\bar{n}$ tetraqurks is the narrowest one ($\sim 10$ MeV). From Table \ref{results-mass-168}, $K_{bc}$ and $K_{ns}$ for this state are both negative while $K_{b\bar{s}}$, $K_{b\bar{n}}$, $K_{c\bar{s}}$, and $K_{c\bar{n}}$ are all positive, which indicates that this tetraquark tends to be a $(qq)(\bar{q}\bar{q})$ but not $(q\bar{q})(q\bar{q})$ structure. Therefore, the narrow rearrangement decay width is consistent with the argument from $K_{ij}$ for this state.

There are six $J^P=1^+$ $bc\bar{s}\bar{n}$ states. The lowest $1^+$ has similar rearrangement decay property to the lowest scalar state: two channels with almost equal coupling matrix elements and comparable widths. The second and third lowest states are almost degenerate, but their dominant coupling channels are different. The former mainly couples to $\bar{B}^0_sD^*$ and $\bar{B}D_s^{*+}$ while the latter to $\bar{B}_s^{*0}D$ and $\bar{B}^*D_s^+$. These channels have important contributions to the decay widths. For the second lowest state, the contribution from the channel $\bar{B}^*D_s^+$ is also significant, although the coupling matrix element is not so large. The third and second highest states are also almost degenerate. Main contributions to the decay width of the third highest state are from the $\bar{B}_s^{*0}D^*$ and $\bar{B}^*D_s^{*+}$ channels as well as the $\bar{B}_s^0D^*$ channel while those to the second highest state are the $\bar{B}^0_sD^*$ and $\bar{B}D_s^{*+}$ channels. As for the highest state around 7570 MeV, it mainly decays into $\bar{B}_s^{*0}D^*$ and $\bar{B}^*D_s^{*+}$. From the signs of $K_{ij}$'s, one cannot get a correct conclusion about the width. From Table \ref{results-mass-168}, only the highest $1^+$ state is probably stable, but the results in Table \ref{results-decay-123} illustrate that the second highest state is relatively stable.

The two $2^+$ states share the same rearrangement decay channels. The lower state has larger width than the higher one does, which cannot be understood just from the signs of $K_{ij}$'s. The reason is that the mass difference between the two states is not large but the lower state has stronger couplings with the two channels.

\subsubsection{$F_2=bs\bar{c}\bar{n}$ and $F_3=bn\bar{c}\bar{s}$ states}

For the four $0^+$ $bs\bar{c}\bar{n}$ states, each state has a dominant decay channel. The second highest one is relatively stable, which is consistent with the argument with $K_{ij}$'s \cite{Wu:2018xdi}. For the three lowest $1^+$ states and the highest $1^+$ state, each one has a dominant decay channel. The second highest $1^+$ $bs\bar{c}\bar{n}$ gets contributions mainly from $\bar{B}^*D_s^{*-}$ and $B_c^-\bar{K}^*$. The third highest state has the dominant decay channel $B_c^-\bar{K}^*$, but the decay into $\bar{B}^*D_s^{*-}$ is also significant. One cannot understand the relatively stable nature of the second lowest state just from $K_{ij}$'s. The broad width for the lowest $1^+$ state is due to the strong coupling with the $B_c^{*-}\bar{K}$ channel and the large phase space. For the two $2^+$ $bs\bar{c}\bar{n}$ tetraquarks, the order of decay width is nothing special. The higher state is broader although the ${cn}$ and ${bs}$ diquarks have slightly attractive color-magnetic interactions.

In the $bn\bar{c}\bar{s}$ case, the features of decay widths and spectrum are similar to the $bs\bar{c}\bar{n}$ case.

\subsection{Systems composed of $b$, $c$, $u$, and $d$ quarks}

Because of the isospin symmetry, there are two systems in this case, $F_4=bu\bar{c}\bar{d}$ and $F_5=bc\bar{u}\bar{d}$. The mass formula we use is
\begin{eqnarray}
M=[M_{X(4140)}-(E_{CMI})_{X(4140)}]+\Delta_{bc}-2\Delta_{sn}+E_{CMI}.
\end{eqnarray}
We have studied the spectrum for the $F_4$ states in Ref. \cite{Wu:2018xdi} and do not repeat the results here. For the $F_5$ tetraquarks, they are also partner states of $T_{cc}$ and their masses with Eq. \eqref{mass-ref} were estimated in Ref. \cite{Luo:2017eub}. Here, with the updated method, we get 100 MeV higher results (still more than 100 MeV lower than the theoretical upper limits), which makes the previously stable lowest $0^+$ and $1^+$ states decay. Compared with the result in Ref. \cite{Karliner:2017qjm} where the lowest $bc\bar{u}\bar{d}$ is slightly below the $\bar{B}D$ threshold, the corresponding state in the present method is slightly above the threshold. If the present CMI model is correct, the $F_4$ and $F_5$ spectra can be related and the observation of one state can be used to predict other states. The effects due to the uncertainties of coupling parameters $C_{ij}$'s on the mass splittings can be understood from the values of $K_{ij}$'s in Table XI of Ref. \cite{Wu:2018xdi} and Table \ref{results-mass-579}.

Now we move on to the decay properties of the $F_4$ states. Since we are interested in tetraquarks with four different flavors, it is not necessary to consider the isoscalar case and we only give results in Table \ref{results-decay-468} for the isovector decay channels. The feature for the four $0^+$ $bu\bar{c}\bar{d}$ is similar to the $F_2=bs\bar{c}\bar{n}$ case where each state has a dominant decay channel. Different from the expectation from $K_{ij}$'s that the second highest state is relatively stable, the second lowest one has a smaller width. The reason is that the state couples mainly to a channel with smaller phase space. For the six $1^+$ $bu\bar{c}\bar{d}$ tetraquarks and the two $2^+$ states, one also finds similar features to the $F_2=bs\bar{c}\bar{n}$ case.

There are two possible isospins for the $F_5=bc\bar{u}\bar{d}$ states. Because of the Pauli principle, the two isospin cases have different decay properties which can be seen from Table \ref{results-decay-579}. The isovector tetraquarks are generally broader than the isoscalar states. The lowest $I=1$ and $I=0$ $0^+$ $bc\bar{u}\bar{d}$ states both decay dominantly to the $\bar{B}D$ channels (note $\bar{B}D$ means both $B^-D^+$ and $\bar{B}^0D^0$). The smaller phase space for the $I=0$ state decides its small width ($\sim5$ MeV with $\alpha=7.27$ GeV). The width of the $I=1$ state is also not very large, although it is about 200 MeV heavier than the $I=0$ state. The higher $I=1$ $0^+$ tetraquark mainly decays to the $\bar{B}^*D^*$ channels. The coupling of the higher $I=1$ $0^+$ state with these channels is also strong, but the decay is kinematically forbidden. The dominant decay channels for the $I=1$ and $I=0$ $1^+$ $bc\bar{u}\bar{d}$ tetraquarks are manifest from Table \ref{results-decay-579}. For the $2^+$ tetraquarks, the width of the isoscalar state is larger than the isovector state, although the former state is lighter. This is because of the stronger coupling matrix element. We cannot judge the stability just from the signs of $K_{ij}$'s, although several states satisfy the condition $K_{qq}<0,K_{q\bar{q}}>0$. If we use $\alpha=7.27$ GeV, the largest decay width for the $F_5$ states is around 35 MeV. Therefore, these possible tetraquark states can be detectable if they do exist. Of course, whether the value of $\alpha$ is reasonable for the double-heavy tetraquarks needs further studies.

\subsection{Systems composed of $b$, $s$, $u$, and $d$ quarks}

Replacing the $c$ quark in $F_4$ and $F_5$ systems with the $s$ quark, one gets the $F_6=bu\bar{s}\bar{d}$ and $F_7=bs\bar{u}\bar{d}$ systems. Now the mass formula reads
\begin{eqnarray}
M=[M_{X(4140)}-(E_{CMI})_{X(4140)}]+\Delta_{bc}-\Delta_{cn}-\Delta_{sn}+E_{CMI}.
\end{eqnarray}
These systems can be generally denoted as $bq\bar{q}\bar{q}$ where $q=u,d,s$. The $X(5568)$ is a $F_6$ state. In Ref. \cite{Liu:2016ogz}, we performed preliminarily a systematic study of the $Qqq\bar{q}$ ($Q=b,c$) states with the CMI model. That study suffered from a large uncertainty due to the quark mass choice and the $SU(3)$ breaking effects. Here, we revisit the systems with the updated scheme. In this subsection, we focus on the $Q=b$ case.

\subsubsection{$F_6=bu\bar{s}\bar{d}$ states}

The mass difference between the lowest and highest tetraquarks is more than 700 MeV. The reason is the system contains three light quarks and the coupling parameters for light quarks are larger than those for heavy quarks. Compared with the former systems, the estimated tetraquark masses of several states will get stronger effects from the uncertainties of coupling parameters, which can be understood from the $K$ factors in Table \ref{results-mass-168}. If the obtained masses are realistic, all the $bu\bar{s}\bar{d}$ tetraquarks have rearrangement channels.

The D0 $X(5568)$ is related to the lowest $0^+$ or $1^+$ $bu\bar{s}\bar{d}$ tetraquark state. Our lowest $0^+$ $bu\bar{s}\bar{d}$ is about 70 MeV higher than the $X(5568)$ mass. It has only one rearrangement decay mode, $\bar{B}_s^0\pi^+$, with $\Gamma\sim$15 MeV. The width is roughly consistent with that of $X(5568)$, $\Gamma=21.9^{+8.1}_{-6.9}$ MeV, but one will get an inconsistent result ($\Gamma\sim$9 MeV) if we adjust the mass to 5568 MeV. The second lowest $bu\bar{s}\bar{d}$ couples mainly to another mode $\bar{B}^0K^+$. Its width is around 20 MeV. For the other two higher $0^+$ states, the dominant decay modes are both $\bar{B}_s^{*0}\rho^+$ and $\bar{B}^{*0}K^{*+}$. Again, one finds that the second highest scalar tetraquark has a relatively stable structure. This is consistent with the argument from $K_{ij}$'s.

What the D0 Collaboration measured is probably also a state around 5616 MeV with $J^P=1^+$. Our lowest $1^+$ $bu\bar{s}\bar{d}$ with $\Gamma\sim 15$ MeV is 80 MeV higher than the experimental result. If we set its mass to 5616 MeV, the width will become a value around 9 MeV.  We still cannot understand the experimental result. For the other $1^+$ states, the third highest one is relatively stable, but one cannot use the $K$ factors to judge its stability. The reason for the relatively narrow width is that it decays mainly into channels having small phase spaces with weaker couplings. For the two $2^+$ states, each one has a dominant rearrangement channel. The higher state is not relatively stable although it satisfies the condition $K_{qq}<0, K_{q\bar{q}}>0$.

\subsubsection{$F_7=bs\bar{u}\bar{d}$ states}

The spectrum has some similar features to the $F_5=bc\bar{u}\bar{d}$ case since the differences come only from the coupling strengths. Compared with the $bc\bar{u}\bar{d}$ tetraquarks, the widths of the low-lying $I=0$ $bs\bar{u}\bar{d}$ are smaller but those of other states are similar.

In Ref. \cite{Yu:2017pmn}, a possible stable $bs\bar{u}\bar{d}$ tetraquark state was proposed by noticing the higher threshold for rearrangement decay. The investigation with a chiral quark model in Ref. \cite{Chen:2018hts} indicates that no stable diquark-antidiquark $bs\bar{u}\bar{d}$ exists but a bound molecule-type state with $I(J^P)=0(0^+)$ is possible. The weak decays of the stable $bs\bar{u}\bar{d}$ were studied in Ref. \cite{Xing:2019hjg}. Our lowest isoscalar $0^+$ and $1^+$ tetraquarks are both slightly below respective thresholds for rearrangement decay channels and they are probably stable states. If their masses are underestimated, their widths should not be larger than those of the corresponding $I=1$ states ($\Gamma\sim 22$ MeV). Therefore, the $bs\bar{u}\bar{d}$ exotic states are worthwhile to search for.

\subsection{Systems composed of $c$, $s$, $u$, and $d$ quarks}

The $F_8=cu\bar{s}\bar{d}$ and $F_9=cs\bar{u}\bar{d}$ systems are obtained with the replacement $b\to c$ for the $F_6$ and $F_7$ systems. The mass formula in this case is
\begin{eqnarray}
M=[M_{X(4140)}-(E_{CMI})_{X(4140)}]-\Delta_{cn}-\Delta_{sn}+E_{CMI}.
\end{eqnarray}
For the spectrum of $F_8$ ($F_9$), the features are similar to those of $F_6$ ($F_7$). However, the decay widths are larger than the corresponding bottom cases. Since the coupling matrix elements have similar values, the difference in width is because of the phase spaces.

In the $F_8$ case, the rearrangement decay width of the lowest $0^+$ $cu\bar{s}\bar{d}$ is about 70 MeV if we use $\alpha=7.27$ GeV. The narrowest tetraquark is the second highest $0^+$ state with $\Gamma\sim 40$ MeV, which is consistent with the argument from the $K$ factors. For the highest and the second lowest $0^+$ tetraquarks, the widths are more than 110 MeV. For the $1^+$ states, the widths range from 62 MeV to 122 MeV. Those for the two $2^+$ states are 54 MeV and 125 MeV. These values are all measurable if our results are realistic.

In the $F_9$ case, the lowest $0^+$ and $1^+$ tetraquarks are both states with narrow widths, decaying strongly with small phase space or decaying weakly. The widths of the isovector tetraquarks (52$\sim$130 MeV) are comparable to those of the $F_8$ states. For the isoscalar states, the higher $0^+$ is not broad ($\Gamma\sim10$ MeV), but the other $1^+$ and $2^+$ states are ($\Gamma=50\sim100$ MeV).

\section{Discussions and summary}\label{sec4}

In studying the properties of the LHCb $P_c$ states \cite{Cheng:2019obk}, we adopted the same CMI model as the present work. In that case, the pentaquark masses estimated with meson-baryon thresholds are consistent with the measured data. However, in the tetraquark case, the masses estimated with meson-meson thresholds may be underestimated, which was illustrated in the $cs\bar{c}\bar{s}$ system \cite{Wu:2016gas}. Thus, we can correct other tetraquark masses by assuming that the $X(4140)$ is a $cs\bar{c}\bar{s}$ tetraquark. With the updated estimation method, the masses based on $X(4140)$ are always larger than those based on meson-meson thresholds and smaller than those according to Eq. \eqref{mass-ori}. The results show that only several states, e.g. lowest $bs\bar{u}\bar{d}$ and $cs\bar{u}\bar{d}$, are probably stable. For most states, strong decays are allowed. It is necessary to investigate their decay widths using appropriate models.

Based on the decay properties of conventional hadrons, one expects several types of two-body strong decays for tetraquarks: 1) rearrangement decays into meson-meson states, 2) baryon-antibaryon decay modes through the creation of a quark-antiquark pair, and 3) decays into lower four-quark states and conventional mesons with a created quark-antiquark pair. Compared with the rearrangement decay patterns, others may be suppressed. In the present study, we only concentrate on the dominant rearrangement decays with a simple model, i.e. $H_{decay}=\alpha$ is a constant. This model has been successfully used to explain the decay ratios for LHCb $P_c$ states \cite{Cheng:2019obk}. The present investigation is also helpful for us to clarify the internal structures of the observed four-quark states through their partial decay widths \cite{Ma:2015nmy}.

To get more information in understanding the nature of exotic states, one should combine the analyses from spectrum and decay properties. For the $X(5568)$ observed by the D0 Collaboration, its flavor structure corresponds to our $0^+$ or $1^+$ $F_6=bu\bar{s}\bar{d}$ states. From the numerical results, one cannot understand its mass and decay width consistently in the tetraquark picture. However, if a narrow $bu\bar{s}\bar{d}$ state about 70 MeV higher were observed, its tetraquark nature could be understood in the present model. On the other hand, the lowest $0^+$ and $1^+$ isoscalar $F_7=bs\bar{u}\bar{d}$ states may be stable although their masses are higher than the lowest $F_6$ state, because the number of decay patterns is reduced for such states which are affected by the Pauli principle.

The lowest $F_9=cs\bar{u}\bar{d}$ tetraquark states, similar to the $F_7$ states are also possibly stable. For the $F_8=cu\bar{s}\bar{d}$ states, we only show results in the $I=1$ case. In fact, the $I=0$ states are degenerate with the $I=1$ states in the present model. According to our results, the lowest $0^+$ isoscalar tetraquark is below the $D_{s0}^*(2317)$ state. On the experimental side, it is interesting to answer whether
a lower exotic $c\bar{s}$-like state exists or not in order to test the model. On the theoretical side, it is worthwhile to investigate the nature of $D_{s0}^*(2317)$ further by considering contributions from both two-quark and four-quark components. Similar studies of $D_{s1}(2460)$ should also be performed.

Another system affected by the Pauli principle is $F_5=bc\bar{u}\bar{d}$. There are controversial results on the masses of the lowest $F_5$ states in the literature. The present method leads to unstable $bc\bar{u}\bar{d}$ tetraquarks. The lowest $0^+$ and $1^+$ isoscalar states are slightly above the $\bar{B}D$ and $\bar{B}^*D$ thresholds, respectively. According to the mass difference, one would have an unstable $1^+$ $T_{cc}=cc\bar{u}\bar{d}$ tetraquark and a stable $1^+$ $T_{bb}=bb\bar{u}\bar{d}$ tetraquark. The predicted widths (several MeVs or around 10 MeV) for the $bc\bar{u}\bar{d}$ states can be tested in future experiments.

Tetraquark states with four different flavors are certainly exotic. The exotic phenomenon is easy to be identified at experiments. However, to distinguish the nature of the observed state, a compact tetraquark or a molecules, is still difficult. Generally speaking, if a stable four-quark state is slightly below a relevant meson-meson threshold, a molecular structure is preferred. If its mass is much lower, a compact tetraquark interpretation may be more natural. To understand its nature further, more studies of decay properties are also essential. If the state does not have strong decay channels, studies of weak decays will be helpful. If it has strong decays, the decay ratios will be helpful.

Another issue one may consider is the production of the tetraquark states. According to the investigation in Ref. \cite{Jin:2016cpv}, the production mechanism of $X(5568)$ at D0 is very strange if it really exists. Up to now, no confirmed multiquark state has been observed in multiproduction processes. Probably this indicates that multiquark states are difficult to produce in such processes \cite{Han:2009jw,Jin:2016cpv}. If this conclusion is correct, maybe one would not observe states studied here in multiproduction processes. However, their production in decays of higher hadrons and in electron-positron annihilation processes could be detectable. The lowest tetraquark states we study are those composed of $c$, $s$, $u$, and $d$ quarks. They can even be produced at BEPCII.

When discussing the stability of tetraquark states, we conjecture that a state is probably relatively stable if $K_{qq}<0$ and $K_{q\bar{q}}>0$. This actually needs further discussions because the rearrangement decay width is determined by the coupling matrix element, the phase space, and the number of decay channels, while the $K$ factors are just directly related to the coupling matrix elements. Only when the $K$ factors can be related to the total width, probably one would get signals about stability. In effect, we find that the above criterion based on $K_{qq}$ and $K_{q\bar{q}}$ is not always reasonable. The reason is that sometimes the rearrangement decays need modified $K$ factors, not those in Eq. \eqref{KijCij}. There is a related problem: is it possible to distinguish molecular structures from compact structures within the CMI model once a four-quark state were observed in its strong decay modes? Notice that the model is based on the hypothesis of the existence of compact multiquark states, resulting directly from the interquark confining forces. This problem demands further analyses. We will consider the above aspects in later works.

To summarize briefly, in the present work, we discuss the spectrum and decay properties of the possible tetraquark states with four different flavors in a color-magnetic interaction model. The tetraquark masses are obtained by estimating the mass difference between the tetraquarks and the $X(4140)$. It is found that most states have rearrangement decays and only lowest $J=0,1$ isoscalar $bs\bar{u}\bar{d}$ and $cs\bar{u}\bar{d}$ are probably stable. From the decay widths calculated with a constant Hamiltonian, the studied tetraquarks should be detectable if they do exist. In particular, the widths of the lowest $I(J^P)=0(0^+,1^+)$ $bc\bar{u}\bar{d}$ tetraquarks are in the order of 10 in units of MeV. From the consistency between the masses and widths of $F_6=bu\bar{s}\bar{d}$ states, it is not natural to interpret the $X(5568)$ as a tetraquark. One of the difficulties encountered by the CMI model is the fact of nonuniversality of the constituent quark masses. When applying the model in the present work, we use the modified mass formula with reference to the mass of $X(4140)$. Although the adopted choice may be subject of discussion, hopefully, the obtained results are helpful for future studies of similar systems.

\section*{Acknowledgements}

This project is supported by National Natural Science Foundation of China (Grants No. 11775130, No. 11775132,
No. 11635009, No. 11325525, No. 11875179) and by Natural Science Foundation of Shandong Province (Grants No. ZR2016AM16, No. ZR2017MA002).

\end{document}